\documentclass[journal,comsoc]{IEEEtran}
\usepackage[T1]{fontenc}
\usepackage[latin9]{inputenc}
\usepackage{color}
\usepackage{array,authblk,amssymb}
\usepackage{multirow}
\usepackage{fixltx2e}
\usepackage{graphicx}
\usepackage{cite}
\usepackage{amsfonts,xcolor,colortbl} 
\usepackage{amsmath}
\usepackage{makecell}
\usepackage{setspace}% http://ctan.org/pkg/setspace
\AtBeginEnvironment{tabular}{\singlespacing \footnotesize}% Single spacing in tabular environment
\usepackage{array}
\newcolumntype{M}[1]{>{\centering\arraybackslash}m{#1}}

\usepackage{amsthm,epstopdf }
\theoremstyle{definition}

\makeatletter
\def\endthebibliography{%
	\def\@noitemerr{\@latex@warning{Empty `thebibliography' environment}}%
	\endlist
}
\makeatother

\makeatother

\begin{document}
	
	\title{Generative AI for Physical Layer Communications: A Survey}
	
	\author{Nguyen Van Huynh, Jiacheng Wang, Hongyang Du, Dinh Thai Hoang, Dusit Niyato, Diep N. Nguyen, Dong In Kim, and Khaled B. Letaief
		
		\thanks{Nguyen Van Huynh is with the School of Computing, Engineering, and the Built Environment, Edinburgh Napier University, Edinburgh EH10 5DT, United Kingdom (e-mail: h.nguyen2@napier.ac.uk)}
		\thanks{Jiacheng Wang, Hongyang Du, and Dusit Niyato are with the School of Computer Science and Engineering, Nanyang Technological University, Singapore 639798 (e-mail: jiacheng.wang@ntu.edu.sg,	hongyang001@e.ntu.edu.sg, dniyato@ntu.edu.sg).}
		\thanks{Dinh Thai Hoang and Diep N. Nguyen are with the School of Electrical and Data Engineering, University of Technology Sydney, Sydney, NSW 2007, Australia (e-mail: hoang.dinh@uts.edu.au, diep.nguyen@uts.edu.au)}
		\thanks{Dong In Kim is with the Department of Electrical and Computer Engineering, Sungkyunkwan University, Suwon 16419, South Korea (e-mail: dikim@skku.ac.kr)}
		\thanks{Khaled B. Letaief  is with the Department of Electrical and Computer Engineering, Hong Kong University of Science and Technology (HKUST), Hong Kong (e-mail: eekhaled@ust.hk)}
	}
	
	\maketitle
	\begin{abstract}
		The recent evolution of generative artificial intelligence (GAI) leads to the emergence of groundbreaking applications such as ChatGPT, which not only enhances the efficiency of digital content production, such as text, audio, video, or even network traffic data, but also enriches its diversity. Beyond digital content creation, GAI's capability in analyzing complex data distributions offers great potential for wireless communications, particularly amidst a rapid expansion of new physical layer communication technologies. For example, the diffusion model can learn input signal distributions and use them to improve the channel estimation accuracy, while the variational autoencoder can model channel distribution and infer latent variables for blind channel equalization. Therefore, this paper presents a comprehensive investigation of GAI's applications for communications at the physical layer, ranging from traditional issues, including signal classification, channel estimation, and equalization, to emerging topics, such as intelligent reflecting surfaces and joint source channel coding. We also compare GAI-enabled physical layer communications with those supported by traditional AI, highlighting GAI's inherent capabilities and unique contributions in these areas. Finally, the paper discusses open issues and proposes several future research directions, laying a foundation for further exploration and advancement of GAI in physical layer communications.
	\end{abstract}
	
	
	%-----------------------------------------------------------------------------------------------------------
	\section{Introduction}\label{sec:intro}
The recent surge in various large-scale datasets, combined with the ongoing progress in artificial intelligence (AI) technologies, has accelerated the development of generative AI (GAI) and led to the creation of GAI based innovative applications like DALL.E and ChatGPT~\cite{ray2023chatgpt}. The emergence of these killer applications has significantly enhanced the efficiency of digital content generation and enriched the variety of the produced content, signifying the arrival of the AI-generated content (AIGC) era~\cite{wang2023guiding}. Unlike traditional AI models, which focus mainly on training, analyzing, and classifying samples, GAI excels in analyzing the distribution characteristics of complex data across different spaces and dimensions, uncovering data patterns~\cite{du2023beyond}. On this basis, GAI can fully utilize the obtained features to generate outputs similar to its input data and present them to users in various forms. A representative example is stableDiffusion~\cite{rombach2022high}, which achieves state-of-the-art scores in class-conditional image synthesis and text-to-image conversion. Different from existing studies focusing on image classification or segmentation, stableDiffusion focuses on the generative, demonstrating greater flexibility and efficiency compared to traditional content creation techniques. Through the fundamental working principles of GAI models and the representative examples, we can see that GAI possesses two core capabilities. The first is the ability to analyze and capture various features of complex data distributions. The second is the utilization of these captured features to generate new data that is similar to, but distinct from, the real data. Therefore, not only does GAI facilitate the generation of digital content, but its potent capability for data distribution analysis also supports research in various domains, including physical layer communications.

In wireless communications, a fundamental role of the physical layer communications involves converting digital data, generated by higher layers of the protocol stack, into a format suitable for transmitting over communication channels. This process encompasses the steps of encoding the data into a bit sequence, modulating these bits onto a carrier wave, and then propagating the modulated signal through the channel. Correspondingly, at the receiver, this layer undertakes the inverse functions, i.e., demodulating the received signal, decoding the bit sequence, and forwarding the data to the higher layers for processing~\cite{liu2021toward}. Beyond these core tasks, the physical layer is entrusted with several other key functions, such as channel access, channel equalization, and multiplexing. Here, the channel access pertains to the process of determining which device is authorized to transmit data over the channel at any particular moment. Equalization involves compensating for the distortion and interference that can occur during transmission over a communication channel. Multiplexing, on the other hand, is the technique of amalgamating multiple data streams into a unified signal for channel transmission. Therefore, the physical layer is integral in shaping the overall reliability, effectiveness, and performance metrics of a wireless communication system~\cite{wang2017deep}.

Given its importance, researchers have conducted in-depth studies on the physical layer, including techniques like beamforming, modulation and demodulation, signal detection, channel estimation, and channel state information (CSI) compression. These techniques are directly linked to the analysis, compression, as well as the feature extraction of complex physical layer data. Conventional research relies on mathematically expressed models. However, in practical applications, the systems could include unknown effects that are almost impossible to be expressed analytically. Therefore, AI models have been introduced to support the physical layer functions of wireless communications. For instance, deep neural networks (DNNs) can learn the relationship between channel inputs and outputs to enhance the accuracy of channel estimation, thereby supporting the physical layer from various perspectives, such as signal detection, channel equalization, and synchronization~\cite{aldossari2019machine}. In addition, deep learning (DL) models were also applied to support the physical layer communications. For example, recurrent neural networks (RNNs) can assist decoding\cite{sattiraju2018performance}, autoencoders can reduce peak-to-average power ratio~\cite{vahdat2020papr}, and Convolutional Neural Networks (CNNs) can compress the CSI in a massive multiple-input multiple-output (MIMO) system~\cite{jo2021adaptive}.

\begin{figure}[!]
	\centering
	\includegraphics[scale=1]{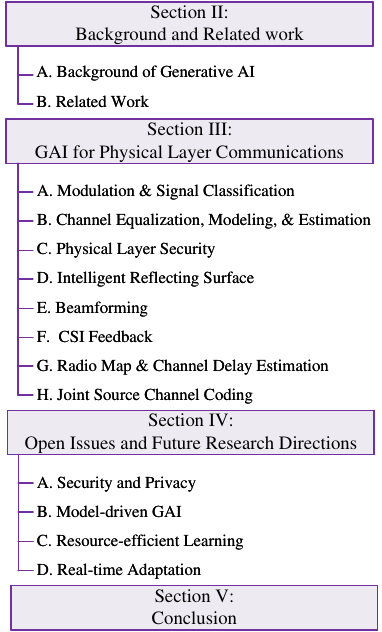}
	\caption{The overall structure of this paper.}
	\label{STCT}
\end{figure}

Although the traditional AI models are effective, their performance is limited. For example, DNNs can learn channel models, but they may struggle or even fail when dealing with channels that are unknown during training. Therefore, researchers introduce GAI, which can not only generate more channel samples to enhance the training data set, but also assist in analyzing the distribution of existing data and extracting its key features, enhancing the system's capability to manage unknown channels~\cite{sun2020generative, ye2020deep}. GAI can also improve physical layer security, beam forming, and other various physical layer techniques. However, applications of GAI have still not been well investigated, especially for emerging technologies such as intelligent reflecting surface (IRS), cell-free, Integrated sensing and communications (ISAC), and extremely large-scale MIMO. Therefore, further advancement of GAI applications in the physical layer communications have been receiving a lot of attentions recently. 

\begin{table*}[!]
	\footnotesize
	\centering
	\caption{\footnotesize LIST OF ABBREVIATIONS} \label{tab:abbreviation} 
	\begin{tabular}{|>{\raggedright\arraybackslash}m{1.6cm}|>{\raggedright\arraybackslash}m{5.7cm}|>{\raggedright\arraybackslash}m{1.6cm}|>{\raggedright\arraybackslash}m{5.7cm}|}
		\hline
		\multicolumn{1}{|>{\centering\arraybackslash}m{1.6cm}|}{\textbf{Abbreviation}} & \multicolumn{1}{>{\centering\arraybackslash}m{5.7cm}|}{\textbf{Description}} & \multicolumn{1}{>{\centering\arraybackslash}m{1.6cm}|}{\textbf{Abbreviation}} & \multicolumn{1}{>{\centering\arraybackslash}m{5.7cm}|}{\textbf{Description}} \\ \hline		\hline
		AI & Artificial Intelligence   & DL & Deep Learning\\ \hline
		TAI & Traditional Artificial Intelligence  & GAI & Generative Artificial Intelligence\\ \hline
		AIGC & AI-generated content  & RNN & Recurrent Neural Network \\ \hline
		DNN &  Deep Neural Network &  CNN & Convolutional Neural Network \\ \hline
		CSI & Channel State Information & ML & Machine Learning  \\ \hline
		SNR & Signal-to-Noise Ratio   & GAN & Generative Adversarial Network \\ \hline
		BER & Bit Error Rate  & PSK  & Phase-Shift Keying \\ \hline
		VAE & Variational Autoencoder & NF & Normalizing Flow \\ \hline
		MIMO & Multiple-Input Multiple-Output & QPSK & Quadrature Phase-Shift Keying \\ \hline
		mmWave & Millimeter Wave & UAV & Unmanned Aerial Vehicle \\ \hline
		NMSE & Normalized Mean Square Error & PLS & Physical Layer Security \\ \hline
		DRL & Deep Reinforcement Learning & RF & Radio Frequency \\ \hline
		IRS & Intelligent Reflecting Surface & BS & Base Station \\ \hline
		UE & User Equipment & FNN & Fully-connected Neural Network \\ \hline
		JSCC & Joint Source Channel Coding & PSNR & Peak Signal-to-Noise Ratio \\ \hline
		AWGN & Additive White Gaussian Noise & WGAN & Wasserstein GAN \\ \hline
		SCMA & Sparse Code Multiple Access & MMSE & Minimum Mean Square Error \\ \hline
	\end{tabular}
\end{table*}

Facing the emerging challenges in the physical layer communications and considering the potential unique support offered by GAI, this paper provides a comprehensive survey of GAI's applications to address diverse problems in physical layer communications. We further discuss comparisons between techniques in the physical layer that are supported by GAI versus those relying on traditional AI models. After that, we discuss the lessons learned from existing studies, emphasizing the key capabilities of GAI employed in these instances. Lastly, the paper highlights open issues and discusses future research directions. The key contributions of this paper are summarized as follows.

\begin{itemize}
	\item From our in-depth investigation, we reveal how to apply different GAI models to solve various physical layer issues. These GAI models encompass not only common ones like generative adversarial networks (GANs) and variational autoencoders (VAEs), but also currently popular diffusion models. Additionally, the physical layer communication issues covered in our survey range from traditional ones like modulation, signal classification, and channel equalization to emerging technologies such as IRS.
	
	\item We examine the problems in physical layer communications supported by traditional AI models and illustrate how physical layer techniques empowered by GAI can address these problems. This reveals the unique support GAI can offer to the physical layer, beyond the capabilities of traditional AI, underscoring the importance of further integrating GAI with physical layer techniques, particularly in dealing with various emerging technologies.
	
	\item We provide an in-depth analysis and summary of the GAI's applications in the physical layer communications, finding that these works primarily leverage three core capabilities of GAI. These include the ability to capture complex data distributions, the capability for cross-dimensional data transformation and processing, and the potential to repair and enhance data. This summary serves as vital guidance for further advancing the applications of GAI in the physical layer.
	
	\item We present significant open issues when applying GAI in the physical layer communications from several perspectives, such as privacy, security, and resource optimization, and provide some directions for future research.
	% \vspace{-0.3cm}
\end{itemize}

The structure of this survey is outlined in Fig.~\ref{STCT}. Section II offers a review of related works, while Section III delves into an in-depth analysis of existing studies. Section IV discusses open issues and future research directions, and Section V concludes the paper. Additionally, Table I lists the abbreviations widely used throughout this survey.
%=============================================
%=============================================

\section{Background and Related Work}\label{sec:Rwork}
This section discusses the background knowledge about GAI and some related surveys, and illustrates the differences between this survey and existing work.
\subsection{Background of Generative AI}
This part introduces the fundamental principles and characteristics of four mainstream GAI models, including 
GANs, VAEs, normalizing flows (NFs), and diffusion models, as these models are frequently utilized in improving physical layer communications.

\begin{figure*}[!]
	\centering
	\includegraphics[scale=0.39]{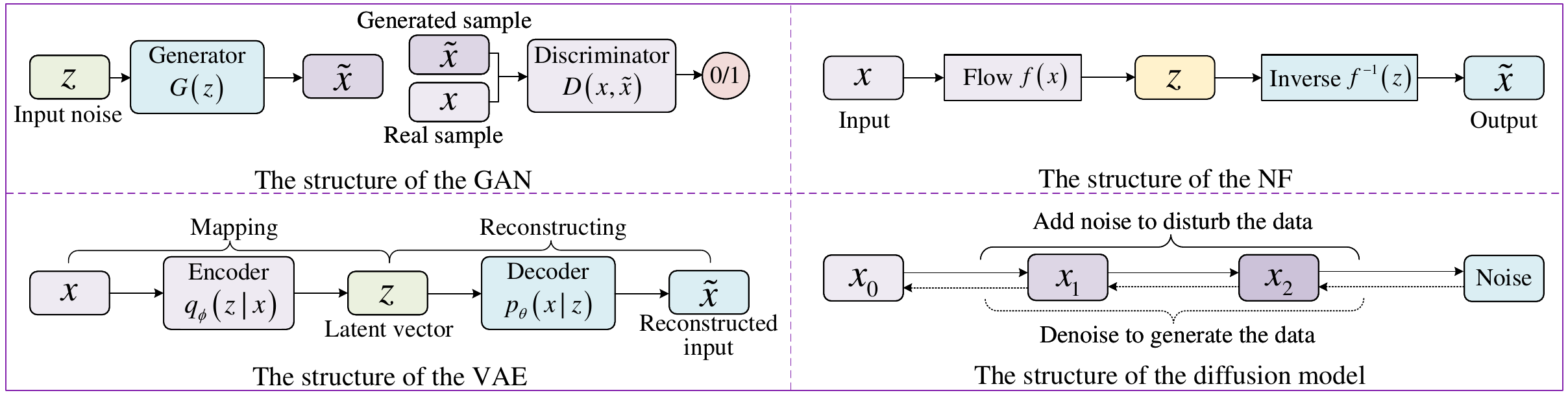}
	\caption{The structure of four GAI models.}
	\label{GAISTCT}
	\vspace{-0.5cm}
\end{figure*}

\subsubsection{\textbf{Generative Adversarial Networks}} A GAN consists of two main elements, i.e., a generator that produces data mimicking real data, and a discriminator that differentiates between the real and generated data. The training process aims for a Nash equilibrium, where the discriminator cannot differentiate between the two~\cite{goodfellow2020generative}. Trained GANs are capable of reconstructing high-dimensional data from low-dimensional input with fewer generator function restrictions compared to other models, which makes them especially proficient in channel estimation~\cite{zhang2021generative} and CSI compression~\cite{tolba2020massive}. Despite these advantages, GANs' training complexity lies in achieving the Nash equilibrium, which is more challenging than optimizing an objective function. This leads to the development of various GAN derivatives, such as StackGAN~\cite{zhang2017stackgan } and PAN~\cite{wang2018perceptual}, focusing either on architecture or objective function optimization~\cite{gui2021review}. These models can be applied across multiple fields like image processing, sequential data handling, and even in drug discovery and malware detection.  

\subsubsection{\textbf{Variational Autoencoders}} VAEs are neural networks designed for compressing and reconstructing data. They differ from traditional autoencoders by using probabilistic methods to model and generate data from a compressed latent space~\cite{doersch2016tutorial}. The VAE comprises an encoder that translates input data into a latent representation, and a decoder that rebuilds the data from this latent space. These components are typically multi-layer neural networks. VAEs optimize their parameters by minimizing a loss function that assesses reconstruction accuracy and aligns the latent space distribution with a prior distribution. Key advantages of VAEs include their ease of implementation and training, effectiveness in learning compressed data representations, and a probabilistic nature that allows for uncertainty estimation and varied outputs~\cite{casale2018gaussian}, thereby providing unique support for signal classification~\cite{zhao2021variational} and joint source-channel coding~\cite{li2023joint}. However, they present challenges in training and parameter tuning, with the possibility of non-interpretable compressed representations.

\subsubsection{\textbf{Normalizing Flows}} NFs are generative models that transform simple probability distributions into complex ones using reversible transformations. Unlike VAEs and GANs, they employ invertible neural networks for these transformations, which include a deterministic mapping function and an adjustable scaling and shifting function~\cite{kobyzev2020normalizing}. The representative examples are the Real NVP~\cite{dinh2016density}, which uses affine coupling layers, and the Masked autoregressive flow~\cite{papamakarios2017masked}, based on autoregressive models. The advantages of NFs lie in efficiently sampling complex distributions, managing high-dimensional data, and learning interpretable latent spaces, which can enhance physical layer techniques, such as signal classification~\cite{he2021learning}. However, the challenges include high computational demands, lengthy training for complex distributions, and transformation function selection. To overcome these, recent studies have explored optimizing architectures and training efficiency through techniques like adversarial training and regularization, demonstrating NFs' potential in diverse applications.

\subsubsection{\textbf{Diffusion Models}} Unlike the above mentioned GAI models, diffusion models start with adding noise to training samples, which is known as the forward diffusion process, and then remove the noise to generate new samples in the inverse process~\cite{yang2022diffusion}. They can be trained on incomplete data in a stable process, enabling them to assist in physical layer technologies like channel modeling~\cite{sengupta2023generative}. However, diffusion models face challenges such as longer sampling times, complex training architectures, and limitations with certain data types~\cite{croitoru2023diffusion}. To address these issues, researchers have developed optimization techniques, such as improving the training speed by reducing variance stochastic gradient descent, adaptive learning rate, and weight normalization.

In Fig.~\ref{GAISTCT}, we present the structures of the four aforementioned GAI models. Additionally, there are other GAI models such as Transformers~\cite{chen2020channel}. These models also possess strong data analysis and modeling capabilities, potentially providing support for physical layer communications.
\subsection{Relate Work}
\subsubsection{\textbf{Generative AI}} Given the GAI's growing popularity, numerous surveys have recently emerged. These surveys focus primarily on the fundamental architecture~\cite{oussidi2018deep, cao2023comprehensive}, principles~\cite{bond2021deep}, implementation methods~\cite{harshvardhan2020comprehensive}, as well as applications~\cite{zhang2023complete,baidoo2023education,de2022deep,qin2023empowering,karapantelakis2023generative,du2023beyond} of GAI models. For instance, the authors in~\cite{cao2023comprehensive} provide a review on the GAI's history, basic components, and recent advances in AIGC across the unimodal and multimodal interactions. Technically,  the authors in~\cite{bond2021deep} present a survey on various deep GAI models, comparing these models, elucidating their underlying principles, interrelations, and reviewing current advancements and applications. For GAI's applications,  the authors in~\cite{du2023beyond} present a practical guide on using GAI for network optimization, demonstrating its effectiveness and contributing to network design. In industry, the authors in~\cite{de2022deep} examine GAI's role in the industrial Internet of Things, focusing on the protection of trust-boundary and the prediction of network traffic, while highlighting challenges to accelerate its adoption. Regarding the emerging Metaverse, the authors in~\cite{qin2023empowering} explore GAI's facilitative role in its development, providing a research roadmap and addressing ethical implications.

\subsubsection{\textbf{AI Enabled Physical Layer Communications}}
% AI models are crucial for advancing physical layer communications, spurring numerous research surveys. The authors in~\cite{qin2019deep} categorize communication systems based on their block structures and explore DL applications in physical layer communications, particularly in signal compression and detection. Another work~\cite{wang2017deep} presents an overview of the autoencoder-based novel communication system, while the authors in~\cite{o2017introduction} treat communication systems as autoencoders and discuss new applications of DL in the physical layer, such as radio transformer networks. About coding, the authors in~\cite{kim2020physical} survey recent advances in DL-based coding, focusing on enhancing the specific coding method using DL techniques. Given the importance of communication delays, authors in~\cite{restuccia2020deep} discuss the need for real-time DL in the physical layer, summarizing the current advancements and limitations in this area. About the security, the authors in~\cite{sharma2023deep} offer a review of DL-based security techniques for addressing issues like attack detection and authentication in 5G and beyond networks. Another work~\cite{kamboj2021machine} reviews three key wireless physical layer security methods, i.e., relay node selection, antenna selection, and authentication, and analyzes their integration with DL. The aforementioned surveys are summarized in Table~\ref{Rwork}. The existing surveys about AI-enabled physical layer technologies and GAI, as discussed, provide two critical insights.
AI models are crucial for advancing physical layer communications, spurring numerous research surveys. These studies primarily concentrate on the application of DL in various domains, including signal detection and compression~\cite{qin2019deep}, coding~\cite{wang2017deep, o2017introduction,kim2020physical}, security~\cite{sharma2023deep,kamboj2021machine}, and communication delay~\cite{restuccia2020deep}. For instance, the authors in~\cite{kim2020physical} survey recent advances in DL-based coding, focusing on enhancing the specific coding method using DL techniques. About the security, the authors in~\cite{sharma2023deep} offer a review of DL-based security techniques for addressing issues like attack detection and authentication in 5G and beyond networks. Given the importance of communication delays, authors in~\cite{restuccia2020deep} discuss the need for real-time DL in the physical layer, summarizing the current advancements and limitations in this area. The aforementioned surveys are summarized in Table~\ref{Rwork}. The existing surveys about AI-enabled physical layer technologies and GAI, as discussed, provide two critical insights.

\begin{table*}[!]
	\renewcommand{\arraystretch}{1}
	\centering
	\caption{Summary of the Related Works}
	\begin{tabular}{|m{0.35cm}|m{1.5cm}|m{13.3cm}|}
		\hline
		\textbf{Ref.} & \textbf{Issue} & \textbf{Key focus of survey }\\
		\hline
		\hline
		\cite{oussidi2018deep} & \multirow{4}{*}{\parbox{1.5cm}{Generative AI}}  & An overview of some GAI models and architectures, training procedures, and limitations of three typical GAI models. \\ \cline{1-1}\cline{3-3}
		\cite{cao2023comprehensive} & & A summary of the history and fundamental components of GAI, along with recent progress in AIGC involving unimodal and multimodal interactions. \\ \cline{1-1}\cline{3-3}
		\cite{bond2021deep} &  & The principles, interrelations, current advancements, and applications of several GAI models. \\ \cline{1-1}\cline{3-3}
		\cite{harshvardhan2020comprehensive} &  & The algorithms and implementation methods of several GAI models, as well as some guidance on selecting GAI models.  \\ \cline{1-1}\cline{3-3}
		\cite{zhang2023complete} &  & Technological development of various AIGC and application of GAI in education and creativity content. \\ \cline{1-1}\cline{3-3}
		\cite{baidoo2023education} &  & An exploration of the advantages and disadvantages of using ChatGPT in educational contexts and some limitations of the ChatGPT. \\ \cline{1-1}\cline{3-3}
		\cite{de2022deep} &  & The state of the art of GAI models and their use in industrial Internet of Things, such as trust-boundary protection, anomaly detection, and so forth. \\ \cline{1-1}\cline{3-3}
		\cite{qin2023empowering} &  & GAI's applications in Metaverse, such as avatars, non-player characters, and virtual world generation, automatic digital twin, and so forth.\\ \cline{1-1}\cline{3-3}
		\cite{karapantelakis2023generative} &  & An extensive overview of recent challenges and developments in applying GAI within mobile communications networks.\\ \cline{1-1}\cline{3-3}
		\cite{du2023beyond} &  & A tutorial of using generative diffusion model in network optimization. \\ \hline\hline
		
		\cite{qin2019deep} & \multirow{10}{*}{\parbox{1.5cm}{AI Enabled Physical Layer Communications}} &  A survey of the recent advancements in DL and its application in signal compression and detection. \\ \cline{1-1}\cline{3-3}
		\cite{wang2017deep} &  & An investigation about the DL-based physical layer, including using DL to redesign the modules in the traditional communication system and replace the communication system with autoencoder-based architecture.     \\ \cline{1-1}\cline{3-3}
		\cite{o2017introduction} &  & Discuss some new applications of DL in the physical layer and present an autoencoder-based physical layer communication system.  \\ \cline{1-1}\cline{3-3}
		\cite{kim2020physical} &  &  An overview of recent advances of DL's applications in coding by focusing on sequential codes and Turbo codes.     \\ \cline{1-1}\cline{3-3}
		\cite{restuccia2020deep} &  & Examine the necessity of real-time DL in the physical layer and provide a summary of the current developments and their limitations. \\ \cline{1-1}\cline{3-3}
		\cite{sharma2023deep} &  &  A detailed examination of different DL and deep reinforcement learning (DRL) methods suited for physical layer security applications.   \\ \cline{1-1}\cline{3-3}
		\cite{kamboj2021machine} &  & The integration of machine learning with the selection of relay nodes, antennas, and authentication processes.   \\ \hline
	\end{tabular}
	\label{Rwork}
\end{table*}

\begin{itemize}
	\item \textit{From the perspective of GAI, the existing surveys primarily discuss the principles, architectures, implementation methods, and the strengths and weaknesses of different mainstream GAI models. Furthermore, researchers analyze the applications of GAI in various domains such as industrial Internet of Things and mobile networks with a variety of applications, and provide future prospects and potential challenges from various aspects, such as ethical impacts and risks.}
	
	\item \textit{Regarding AI-enabled physical layer communications, existing works review the physical layer technologies and various DL techniques. Besides, they present a detailed discussion of how DL supports various physical layer technologies, including signal compression and detection, coding theory, attack detection, physical layer authentication, and so forth.}
\end{itemize}

Despite the comprehensiveness of these surveys, a gap remains in exploring GAI's applications in the physical layer communications. Given the challenges posed by emerging technologies to the physical layer and the unique potential of GAI, this paper delves into how GAI underpins physical layer technologies. We further enhance this exploration by contrasting GAI-assisted physical layer technologies with those reliant on traditional AI models, thereby addressing current research gaps and providing insights into the ongoing evolution of GAI in physical layer communications.
%=============================================
%=============================================
	\section{Generative AI for Physical Layer Communications}\label{sec:survey}
In this section, we provide a comprehensive review of various applications of GAI for physical layer communications. In particular, we first highlight GAI-based approaches for modulation recognition and signal classification. Then, the applications of GAI for channel modeling and estimation, physical layer security, beamforming, and joint source channel coding (JSCC) are discussed in detail. Finally, we review applications of GAI for emerging problems in wireless communications, including IRS, CSI feedback, and radio map estimation.

\subsection{Modulation and Signal Classification}
Signal classification and modulation recognition are always among the most important components in designing receivers of wireless communication systems~\cite{xu2020wavelet}. Specifically, the goal of radio signal classification is to accurately recover information transmitted from the transmitter over a noisy wireless channel by analyzing received signals based on different techniques, such as maximum-likelihood and minimum mean square error (MMSE). On the other hand, modulation recognition aims to detect the modulation technique used at the transmitter and obtain the original transmitted signals by using the corresponding demodulation approach. Signal classification and modulation recognition have been extensively studied and developed since the creation of wireless communications. However, traditional approaches like maximum-likelihood and MMSE require a specific sophisticated mathematical model for each type of wireless channels and environments~\cite{wang2017deep}. In addition, perfect or highly accurate knowledge of the underlying channel and CSI are usually required to obtain good detection performance~\cite{sun2020generative}. However, these approaches appear to be ineffective in future wireless communication systems due to the increased complexity of signals, spectrum efficiency requirements, and the dynamics and uncertainty of UEs' behaviors and characteristics.

To overcome these challenges, DL is emerging as a prominent solution. This is because DL-based approaches can leverage DNNs to learn the relationship between the channel inputs and channel outputs, resulting in data-driven signal detection without requiring any knowledge of channel models. Unfortunately, DL-based solutions require large datasets and long training time to obtain good detection performance~\cite{liu2020deep, van2022transfer}, especially when channel environments change fast due to the user mobility. In practice, collecting and processing enough training data are costly, time-consuming, and sometimes impossible. Moreover, DL-based approaches cannot effectively deal with the dynamics and uncertainty of wireless communications. In particular, a trained DL model only works well with some specific wireless environments that have similar characteristics to the trained environment. In new wireless environments with different conditions, e.g., channel models, surrounding interference, and noise distributions, this trained model will need to be retrained with a huge volume of new training data, which may not be feasible in practice. In addition, conventional DL-based solutions are less effective in modeling complex wireless channels that are time-varying, non-i.i.d distributed, or non-differentiable~\cite{sun2020generative, ye2020deep}. To deal with these limitations and facilitate applications of DL in modulation and signal classification, GAI, with its great capabilities to understand, capture, and generate the distribution of complex and high-dimensional data~\cite{du2023age, wen2023generative}, is a promising approach, as summarized in Table~\ref{tab:Summary_modulation_signal_classification}.

Specifically, the authors in~\cite{sun2020generative} point out that traditional DL-based approaches do not perform well with non-Gaussian and time-varying channels, especially in the low signal-to-noise ratio (SNR) regions. For that, they propose a novel GAN to help the receiver intelligently adapt to the dynamics of wireless channels without retraining DNNs. In particular, the proposed GAN is used to efficiently learn the channel transition probability, i.e., the likelihood function. Then, the estimated channel transition probability is fed into the Viterbi algorithm~\cite{shlezinger2020viterbinet} to derive the maximum-likelihood sequence detection. Moreover, the authors develop an online adjustment policy to fine-tune the proposed GAN network by leveraging the soft output of the model as well as pilot signals, making it more effective with time-varying wireless channels. The numerical results then demonstrate that the proposed GAN network can achieve a bit error rate (BER) of $10^{-2}$ at 8 dB SNR while the ViterbiNet approach~\cite{shlezinger2020viterbinet} can only obtain this level of BER at 12 dB SNR. Moreover, the authors show that by using GAN they can obtain near-optimal BER performance under dynamic channel conditions.

\begin{figure}[!]
	\centering
	\includegraphics[scale=0.27]{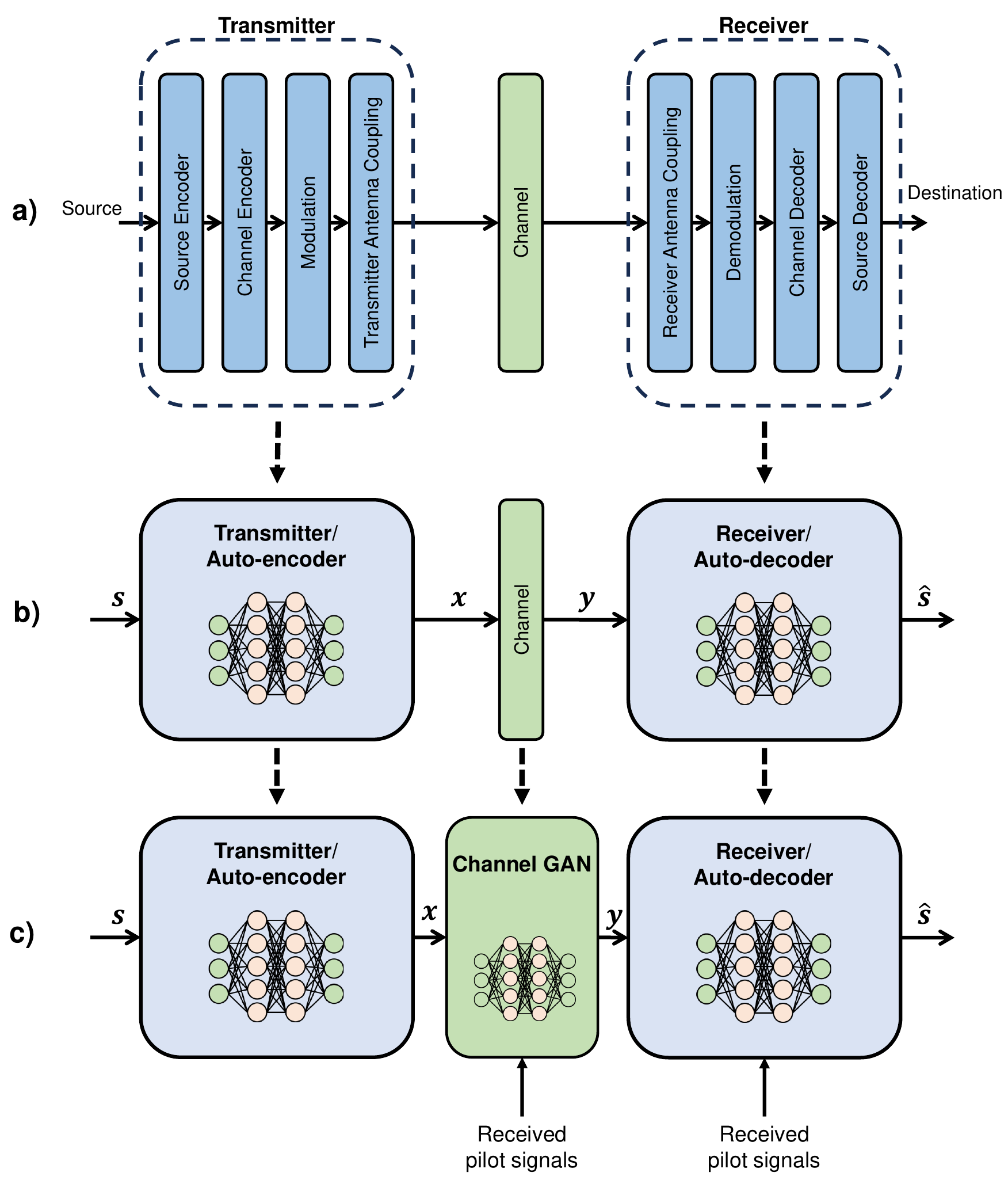}
	\caption{Architectures of traditional wireless communication systems and end-to-end learning-based communication systems: (a) traditional wireless systems, (b) end-to-end communication systems based on autoencoder, where the transceivers are represented by DNNs, and (c) end-to-end communication systems with channel GAN~\cite{ye2020deep}.}
	\label{Fig.GAN_signal_classfication}
\end{figure}

Considering the same GAI method, the authors in~\cite{ye2020deep} also develop a GAN network to model unknown channels in end-to-end wireless communication systems. As depicted in Fig.~\ref{Fig.GAN_signal_classfication}, the authors first consider an end-to-end communication system in which all the signal processing blocks at the transceivers are replaced by DNNs to jointly optimize the performance of the whole system. To do that, traditional DL-based approaches usually assume the availability of CSI and prior channel knowledge which are not always available in practice. To tackle this challenge, the authors design a novel conditional GAN network to represent the channel between the transmitter and the receiver to allow the gradient from the receiver to back-propagate to the transmitter. Moreover, the pilot signals received at the receiver are used as the conditional information of the proposed GAN network, as illustrated in Fig.~\ref{Fig.GAN_signal_classfication}(c). In this way, the GAN network can generate more realistic coefficients for time-varying channels, and thus the end-to-end loss can be optimized to minimize the BER of the system. Interestingly, the authors demonstrate that the Kullback-Leibler divergence of the proposed GAN network can be significantly reduced when training the model over a long period, indicating that the generated data's distribution converges to the target distribution. One potential research direction is to test the proposed architecture in real wireless scenarios to evaluate its effectiveness in dealing with various imperfections of wireless channels.

Besides signal classification, GAN can also be adopted for modulation recognition. For instance, the authors in~\cite{tang2018digital} highlight that an application of DL for signal modulation recognition is often hindered by insufficient training data and overfitting. As such, the authors propose an auxiliary classifier GAN to enlarge the training dataset by generating new data while maintaining high-level features learned from the original training data. The authors then demonstrate that the proposed GAN solution can increase the classification accuracy by up to 6\% compared to conventional DL-based solutions, e.g., AlexNet. Similarly, the authors in~\cite{lee2023generative} propose a GAN network to restore missing signals due to errors in dynamic spectrum sensing or signal sensing. The experimental results then show that the proposed GAN network can preserve each modulation type's global structure and restore up to 50\% missing samples in the time domain. Differently, the authors in~\cite{shtaiwi2022mixture} propose a GAN-based modulation classification approach that is resilient to adversarial attacks. Specifically, the authors indicate that conventional DL-based automatic modulation recognition methods are vulnerable to adversarial attacks with well-designed perturbation injected into wireless channels. To tackle this practical issue, the authors propose a novel GAN network to generate plausible samples that are similar to the received frames. The generated frames are then compared with the perturbed received signals to detect the true class of the modulated signals. The authors then revealed that using only one generator may face the mode collapse problem when dealing with multiple modulation types. As such, multiple generators are incorporated into the proposed GAN network with each generator being used to deal with a specific type of modulation. In this way, the proposed solution can work well with various modulation types. Simulation results show that the proposed GAN model can significantly increase the accuracy of DL-based modulation recognition methods under adversarial attacks. For example, the recognition accuracy for 8 phase-shift keying (PSK) scenarios can be increased from 9\% to around 70\% by using the proposed GAN model.

\begin{figure}[h]
	\centering
	\includegraphics[scale=0.25]{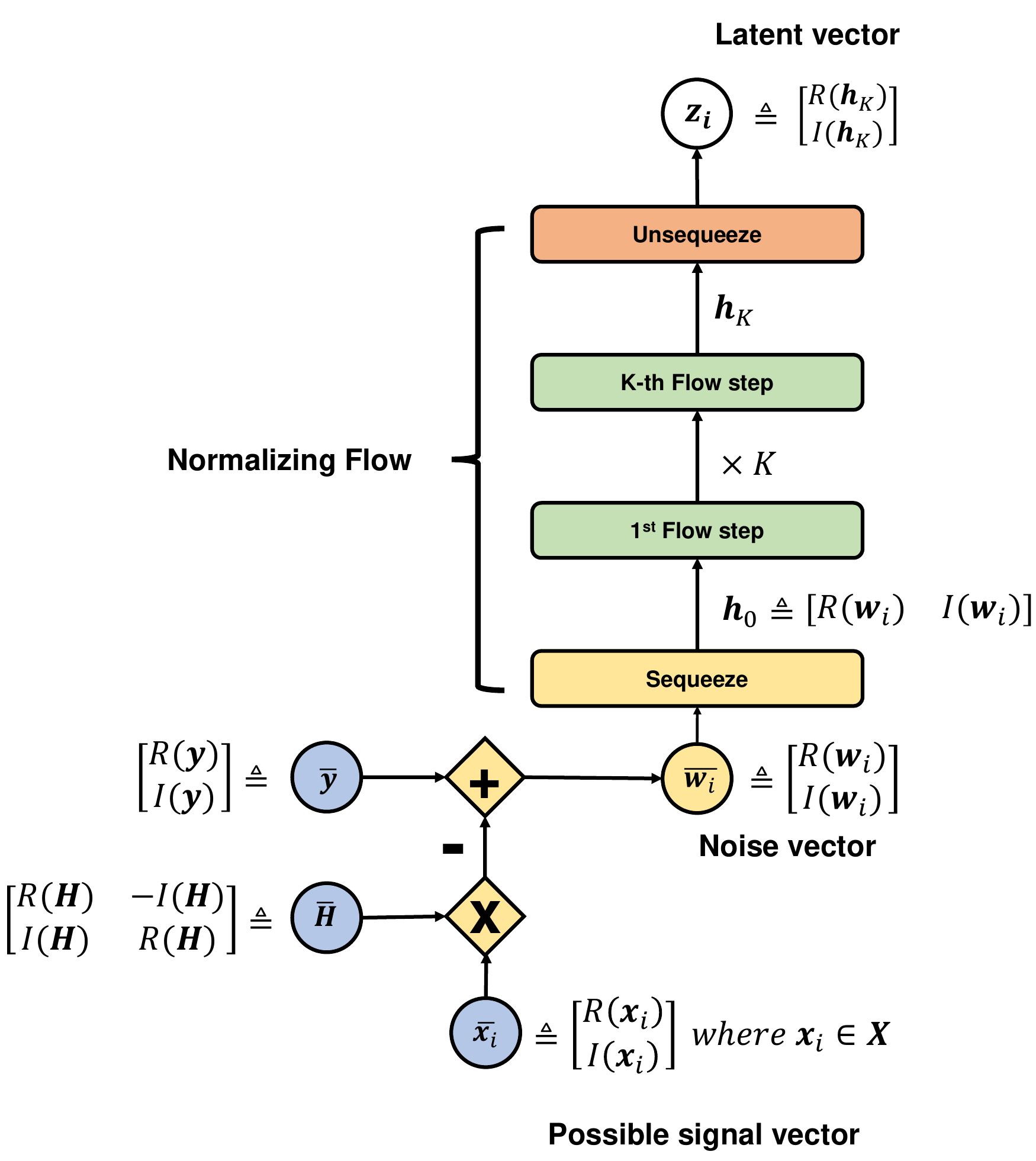}
	\caption{Architecture of the detection framework with NFs consisting of three main components: (i) a squeeze layer, (ii) $K$ flow steps, and (iii) an unsqueeze layer. The noise vector for each possible signal vector is first calculated and fed into the NF. After that, the output of the NF is mapped into the latent space $\mathbf{z}_i$~\cite{he2021learning}.}
	\label{Fig.normalizing_flow_signal_classfication}
\end{figure}

\begin{table*}[!]
	\centering
	\caption{Summary of GAI Approaches for Modulation Recognition and Signal Classification}
	\begin{tabular}{|M{0.6cm}|m{1.5cm}|m{6cm}|m{6.6cm}|}
		\hline
		\textbf{No.} & \textbf{Problem} & \textbf{Drawbacks of TAI} & \textbf{Proposed Approach}\\
		\hline
		\hline
		\cite{tang2018digital} & \multirow{4}{*}{\parbox{1.5cm}{Modulation Recognition}}  & Poor performance due to insufficient data and overfitting &   Using the auxiliary classifier GANs to enlarge training datasets     \\ \cline{1-1}\cline{3-4}
		\cite{li2018generative} & & Require large numbers of labeled samples &   Use GAN to generate samples from noise and labeled data \\ \cline{1-1}\cline{3-4}
		\cite{shtaiwi2022mixture} &  & Vulnerable to adversarial attacks with well-designed perturbation &  Propose a novel GAN network that consists of four generators to improve the model's accuracy and robustness against adversarial attacks   \\ \cline{1-1}\cline{3-4}
		\cite{lee2023generative} &  & Require a large amount of training data, and the training accuracy can be greatly affected by the training data's quality &  Propose a GAN-based method to generate missing wireless signal samples  \\ \hline\hline
		
		\cite{zhao2018application} & \multirow{10}{*}{\parbox{1.5cm}{Signal Classification}} & Lack of clean training dataset. Information loss during the feature extraction process. &   Use GAN to generate a large training dataset without requiring manual annotation. Discriminative model can improve the signal classification process    \\ \cline{1-1}\cline{3-4}
		\cite{ye2020deep} &  & Poor performance due to the dynamics of wireless channels   & Propose a conditional GAN to represent channel effects  \\ \cline{1-1}\cline{3-4}
		\cite{zhao2021variational} &  & Not effective in estimating posterior distributions &   Use VAE to approximate intractable posterior distributions  \\ \cline{1-1}\cline{3-4}
		\cite{li2021variational} &  &  Enormous data labels are required  &   Use VAE to simplify the maximum-likelihood estimation which contains latent variables   \\ \cline{1-1}\cline{3-4}
		\cite{sun2020generative} &  & Require a huge volume of training data. Not perform well when the underlying channel models are completely unknown   &   Use GAN to directly approximate the transition probability of the underlying wireless channel  \\ \cline{1-1}\cline{3-4}
		\cite{he2021learning} &  &  Performance is not guaranteed when the noise statistics is unknown  &  Leverage a NF to effectively learn the distribution of unknown noise  \\ \cline{1-1}\cline{3-4}
		\cite{alawad2022new} &  &  Require more bandwidth resources  &   Use VAE as a probabilistic model to recover transmitted symbols   \\ \cline{1-1}\cline{3-4}
		\cite{alawad2022innovative} &  & Sub-optimal in certain cases   &  Use VAE as a probabilistic model to recover transmitted short-packet symbols   \\ \hline\hline
	\end{tabular}
	\label{tab:Summary_modulation_signal_classification}
\end{table*}

While most existing works in the literature adopt GAN, VAEs and NFs have been gaining attention recently~\cite{li2021variational, zhao2021variational, alawad2022new, he2021learning} due to their capabilities in dealing with signals in the time domain. For example, the authors in~\cite{zhao2021variational} consider the signal classification problem in MIMO orthogonal frequency division multiplexing with index modulation systems. In particular, due to the high complexity of calculating the posterior probability, the authors estimate the variational posterior probability by performing variational inference in a VAE network. Practically, the proposed VAE network can approximate intractable posterior distributions by training an encoder to map input data to a latent distribution and training a decoder to estimate the inputs, making it more effective than conventional DL approaches in classifying received signals when approximating the posterior distribution is complex. Simulation results then reveal that the proposed approach can obtain near-optimal maximum-likelihood performance under different single antenna settings. By using the NF technique, the authors in~\cite{he2021learning} propose a novel signal detection framework, which is fully probabilistic, to approximate unknown noise distributions. Specifically, the authors consider the signal detection problem in MIMO systems with unknown statistical knowledge of noise which is very challenging for traditional DL-based approaches. The authors then utilize the NF technique to design a flexible detection framework that does not require any noise statistics as depicted in Fig.~\ref{Fig.normalizing_flow_signal_classfication}. The proposed NF is constructed by three major components, including an unsqueeze layer, $K$ flow steps, and a squeeze layer. To obtain the maximum-likelihood estimation, the authors first calculate the noise vector $\mathbf{w}_i = \mathbf{y} - \mathbf{H}\mathbf{x}_i$, corresponding to signal vector $\mathbf{x}_i$ with received signal $\mathbf{y}$ and channel matrix $\mathbf{H}$. The proposed NF then maps $\mathbf{w}_i$ into the latent space which consists of latent variable $\mathbf{z}_i$ and the log-determinant. In this way, the corresponding likelihood $p(\mathbf{y}|\mathbf{x_i})$ can be calculated, resulting in accurate maximum log-likelihood estimation. Extensive simulations demonstrate that the proposed framework outperforms existing DL-based methods in terms of BER under non-analytical noise settings. For example, in the quadrature phase-shift keying (QPSK) modulated 4 $\times$ 4 MIMO system, the proposed approach can reduce the detection error of the DetNet architecture~\cite{samuel2019learning} by 39.61\% with SNR=25 dB. However, the performance gap between the proposed method and the traditional maximum-likelihood approach is still noticeable. One potential solution is leveraging the auto-distribution technique to further improve the convergence of the proposed method in unknown noise conditions.

%\cite{zhao2018application,tang2018digital,li2018generative,zhou2020wireless,ye2020deep,zhao2021variational,li2021variational,sun2020generative,he2021learning, alawad2022new,shtaiwi2022mixture,alawad2022innovative,lee2023generative}

\subsection{Channel Equalization, Modeling, and Estimation}

In wireless communications, channel equalization, modeling, and estimation play essential roles in helping the receiver detect the received signals more efficiently. In particular, channel equalization refers to the process of compensating for the distortions incurred when transmitting signals through the communication channel. On the other hand, channel modeling and channel estimation aim to create a mathematical model for the communication channel and estimate the parameters of the channel model, respectively. Over the past few years, DL has been widely adopted for channel equalization, modeling, and estimation both in academia and industry~\cite{ye2017initial, aldossari2019machine}. Unfortunately, DL-based approaches require a huge volume of labeled data to learn sufficient characteristics of a specific channel, and thus limiting their application in dynamic wireless environments with high levels of randomness and variability. In addition, standard neural networks can work well for discriminative tasks but perform poorly when modeling the full complexity of channel distributions. Finally, conventional DL-based methods use a general loss function that makes their predictions less accurate, especially in low SNR regions~\cite{ye2022channel}. For that, GAI has been adopted widely recently for equalizing, modeling, and estimating wireless channels, as summarized in Table~\ref{tab:Summary_channel_estimation_equalization}. Compared with conventional DL techniques, GAI possesses several advantages. Specifically, GAI can generate synthetic training data that is similar to the data it was trained on for ML models of channel estimation. In addition, it can generate data following specific constraints or conditions and leverage data from a source system to generate training data for a target system. All these special features make GAI an ideal tool for channel modeling. Moreover, GAI can be used as an equalizer to learn the mapping from distorted signals to transmitted signals as well as to model the posterior distribution of transmitted signals and then estimate clean signals from distorted observations at the receiver.

\subsubsection{Channel Equalization}

\begin{figure*}[!]
	\centering
	\includegraphics[scale=0.31]{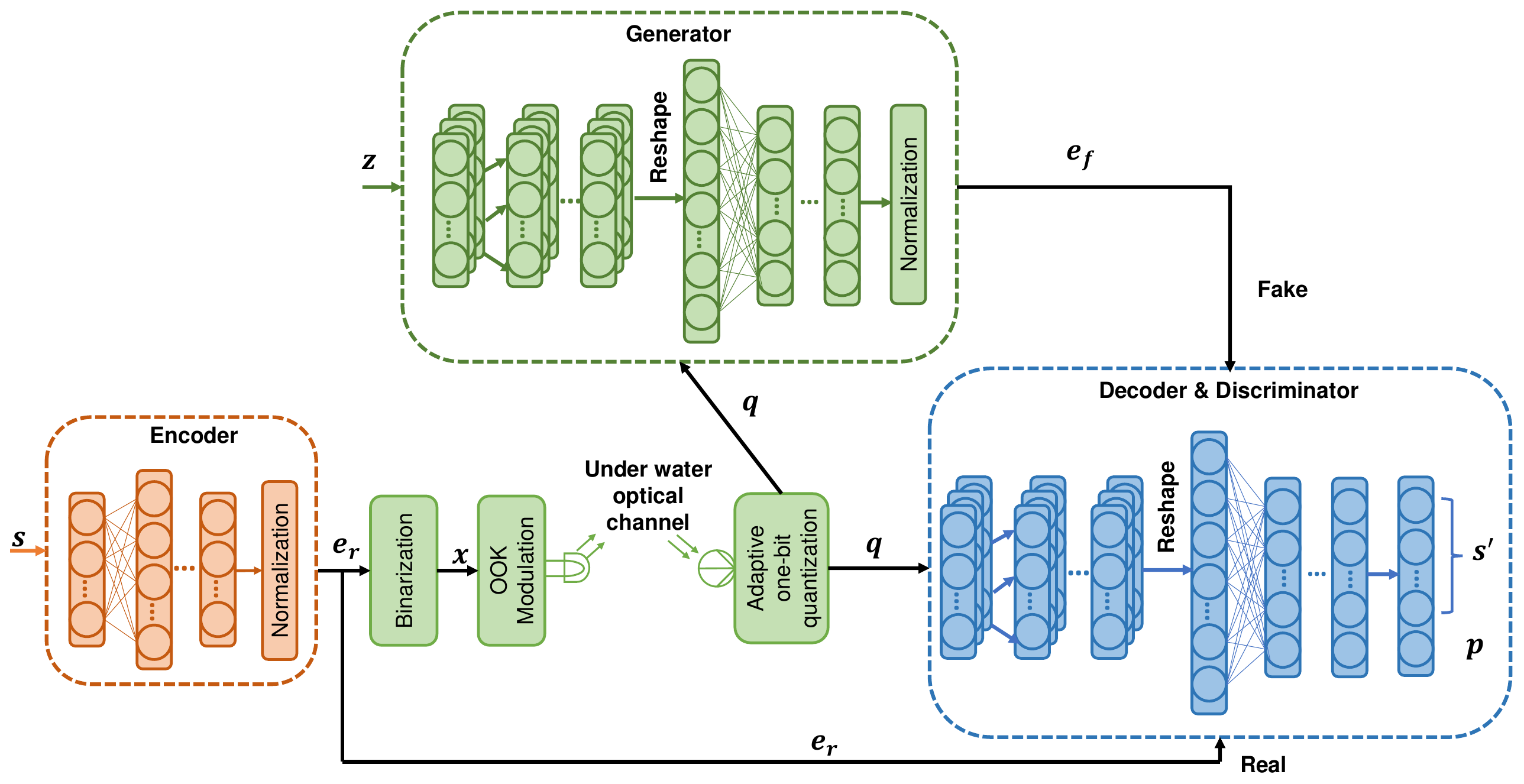}
	\caption{Training structure of the proposed hybrid AE-GAN in which the encoder encodes input signal $\mathbf{s}$, the generator learns the distribution of the real encoded signal $\mathbf{e}_\mathrm{r}$ with the quantized signal $\mathbf{q}$ as its input, and finally the discriminator distinguishes the real encoded signal $\mathbf{e}_\mathrm{r}$ and the fake encoded signal $\mathbf{e}_\mathrm{f}$ generated by the generator~\cite{zou2023underwater}.}
	\label{Fig.GAN_underwater_channel_equalization}
\end{figure*}

In~\cite{zou2023underwater}, the authors develop a hybrid GAN and autoencoder approach for channel equalization of underwater wireless communications with one-bit quantization. Specifically, it is highlighted that underwater wireless communications are extremely vulnerable to severe channel fading caused by the scattering and absorption of underwater environments. Moreover, the strong nonlinearities of one-bit quantization can greatly affect communication reliability. Given these challenges, using conventional DL-based approaches, e.g., autoencoder, may not yield good communication performance. For that, the authors propose to integrate GAN into their autoencoder architecture to significantly improve the channel equalization performance as illustrated in Fig.~\ref{Fig.GAN_underwater_channel_equalization}. Specifically, input signal $\mathbf{s}$ is first encoded by the encoder and then quantized by the adaptive one-bit analog-to-digital converter to reduce the energy consumption of the receiver. The generator of the proposed GAN architecture is used to approximate the distribution of encoded signal $\mathbf{e}_\mathrm{r}$ given quantized signal $\mathbf{q}$ as its input. The discriminator then can distinguish the real encoded signal $\mathbf{e}_\mathrm{r}$ and synthetic encoded signal $\mathbf{e}_\mathrm{f}$ produced by the generator. Finally, the decoder will be used to recover the transmitted signal. In this way, the authors can construct a generalized channel equalization to equalize the one-bit quantization's distortion as well as the severe channel fading of underwater environments.

Due to its capabilities in analyzing signals in the time domain, VAEs have been widely adopted for channel equalization recently. For example, the authors in~\cite{caciularu2018blind} propose to use VAEs for blind channel equalization which is challenging for conventional AI approaches. In particular, an encoder is used to represent the channel model and noise, and a decoder is used to approximate the posterior distribution of transmitted symbols from the received signals. In this way, VAEs can effectively model complex channel distributions and perform inference on latent variables. The authors then extend this work and propose a VAE equalizer for noisy linear and nonlinear channels in~\cite{caciularu2020unsupervised}. They demonstrate that the proposed VAE equalizer significantly outperforms baseline blind equalizers and can obtain similar performance to that of a non-blind adaptive linear MMSE equalizer while not requiring prior knowledge of impulse responses as well as pilot signals. Differently, the authors in~\cite{wu2023cddm} propose to use a diffusion model to remove channel noise. In particular, the proposed channel denoising diffusion model is added as a new physical layer module right after the channel equalization to learn the input signals' distributions and then leverage them to further remove the channel noise. Experiments demonstrate that the proposed diffusion model can significantly reduce the mean square error and outperform existing approaches. For example, at SNR=20 dB under Rayleigh fading, the proposed diffusion model can achieve a 1.06 dB gain compared to the joint source-channel coding system.

\subsubsection{Channel Modeling}

GAI also finds its applications in channel modeling~\cite{hu2022multi, sengupta2023generative, zhang2021distributed}. In~\cite{hu2022multi}, the authors propose to use GAN to model millimeter wave (mmWave) channels. They highlight that accurately modeling mmWave channels is challenging due to several factors such as multiple high frequencies and highly directional beams. For that, the authors design a GAN approach to generate random profiles that include all information about the channel, including channel gains, delays, angle of arrival, and angle of departure of all links between the receiver and the transmitter. Simulation results then show that by using GAN, the authors can generate new channel data that have almost the same cumulative distribution function as real data. With this newly generated data, the authors then can effectively model mmWave channels by capturing the joint distribution of all links between the transmitter and the receiver with multiple frequencies. Similarly, the authors in~\cite{sengupta2023generative} also address the lack of data problem in modeling wireless channels by proposing a diffusion model. The proposed diffusion model can learn to generate new data samples by iteratively adding noise to the previous input, resulting in a more stable training process compared to existing GAN approaches, as demonstrated by the authors.

In~\cite{zhang2021distributed}, the authors introduce a distributed GAN approach to model mmWave channels in unmanned aerial vehicle (UAV) networks. In particular, the authors state that existing approaches for channel modeling using conventional AI as well as centralized GAN are limited by the lack of training channel samples and environmental measurements. For that, they propose to use UAVs to collect mmWave channel data during their aerial services. Each UAV employs GAN to train a local channel model. After that, the generated channel samples produced from the local channel model will be shared with other UAVs in the networks to improve their training process. Extensive simulations show that the proposed distributed GAN approach can significantly improve the modeling accuracy as well as increase the communication rate by 10\% under real-time channel estimation compared to standalone training.

\begin{figure}[!]
	\centering
	\includegraphics[scale=0.16]{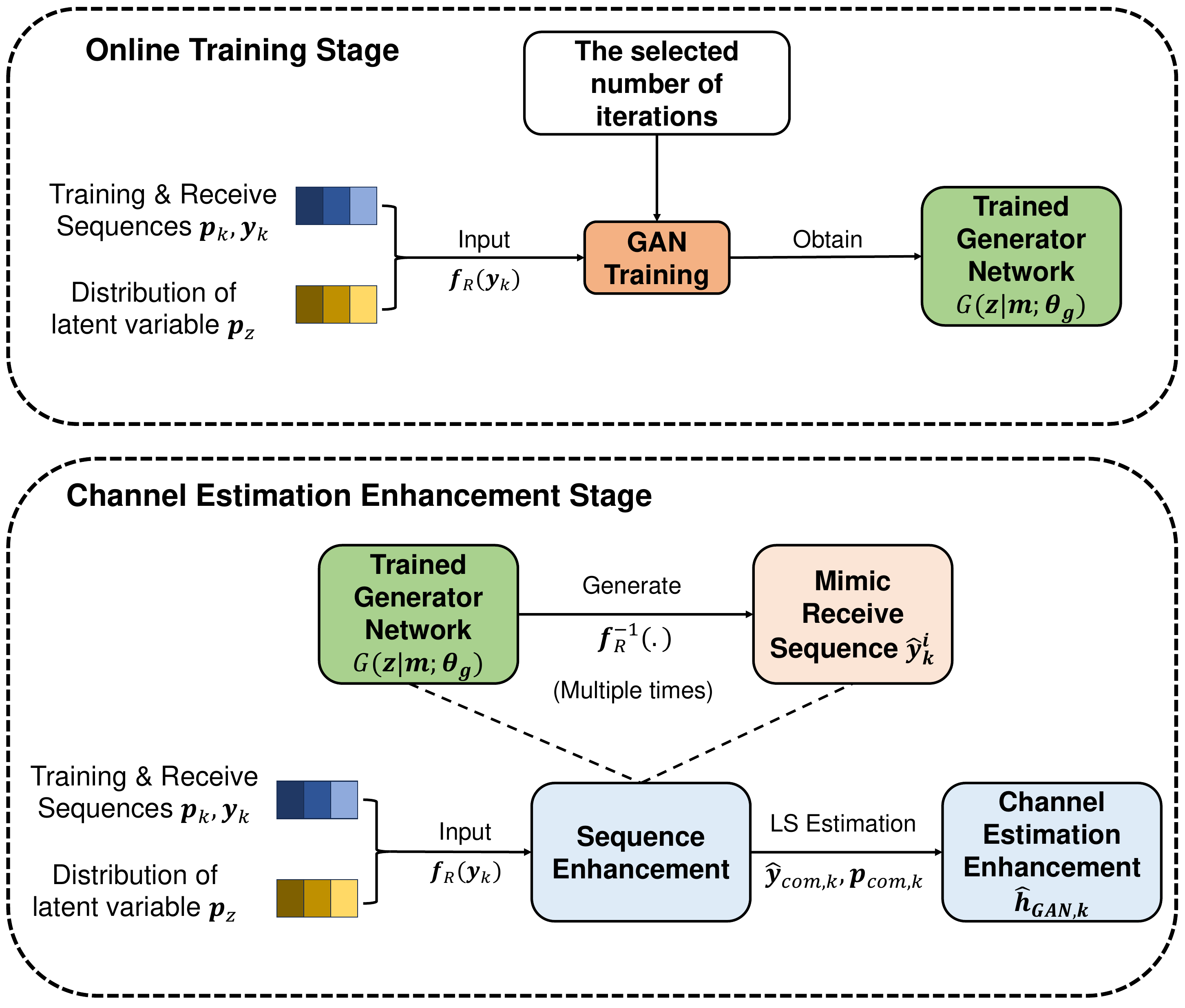}
	\caption{The flowchart of the proposed GAN-based channel estimation. The training sequence $\mathbf{p}_k$ and the receive sequence $\mathbf{y}_k$ are used as the input data of the GAN architecture. The generator then ``mimics'' receive sequences $\hat{\mathbf{y}}_k$ matching with the distribution of the true channel~\cite{hu2020channel}.}
	\label{Fig.GAN_online_training_phase}
\end{figure}

\subsubsection{Channel Estimation}

Besides channel equalization and channel modeling, GAI has been widely adopted in the literature for channel estimation. For example, the authors in~\cite{balevi2021wideband} propose a GAN architecture for wideband channel estimation in mmWave and THz communications. The authors highlight that DL has been widely adopted for channel estimation in recent years. However, conventional DL-based approaches require long pilot sequences to achieve good estimation performance. Moreover, they provide poor channel estimation performance under high channel correlations and high propagation losses. For that, the authors propose to use GAN to estimate frequency selective channels at low SNR regions with short pilot sequences. Specifically, the proposed GAN approach can learn to generate realistic channel coefficients based on a real-world but unknown channel distribution during the offline training phase. After that, the trained GAN network is used as a prior model for online channel estimation by optimizing the input vector of the model based on the current signal received at the receiver. By doing this, the proposed GAN approach can obtain higher channel estimation accuracy with 70\% fewer pilots compared to the traditional CNN networks (e.g., ResNet). Interestingly, the proposed solution can work well when changing the environment's factors such as the number of rays and clusters without retraining the GAN network. Differently, the authors in~\cite{hu2020channel} adopt GAN during the online training phase to further improve the channel estimation performance as illustrated in Fig.~\ref{Fig.GAN_online_training_phase}. Specifically, the receive sequence $\mathbf{y}_k$ and the training sequence $\mathbf{p}_k$ are fed into the proposed GAN architecture as its input data for training. The generator network then can ``mimic'' receive sequences $\hat{\mathbf{y}}_k$ matching with the distribution of the true channel. After that, a newly proposed enhancement algorithm will perform channel estimation based on these new receive sequences. Simulation results indicate that the proposed GAN approach can help to improve the estimation accuracy of traditional training-based channel estimation approaches, especially at low SNRs.

\begin{figure}[!]
	\centering
	\includegraphics[scale=0.15]{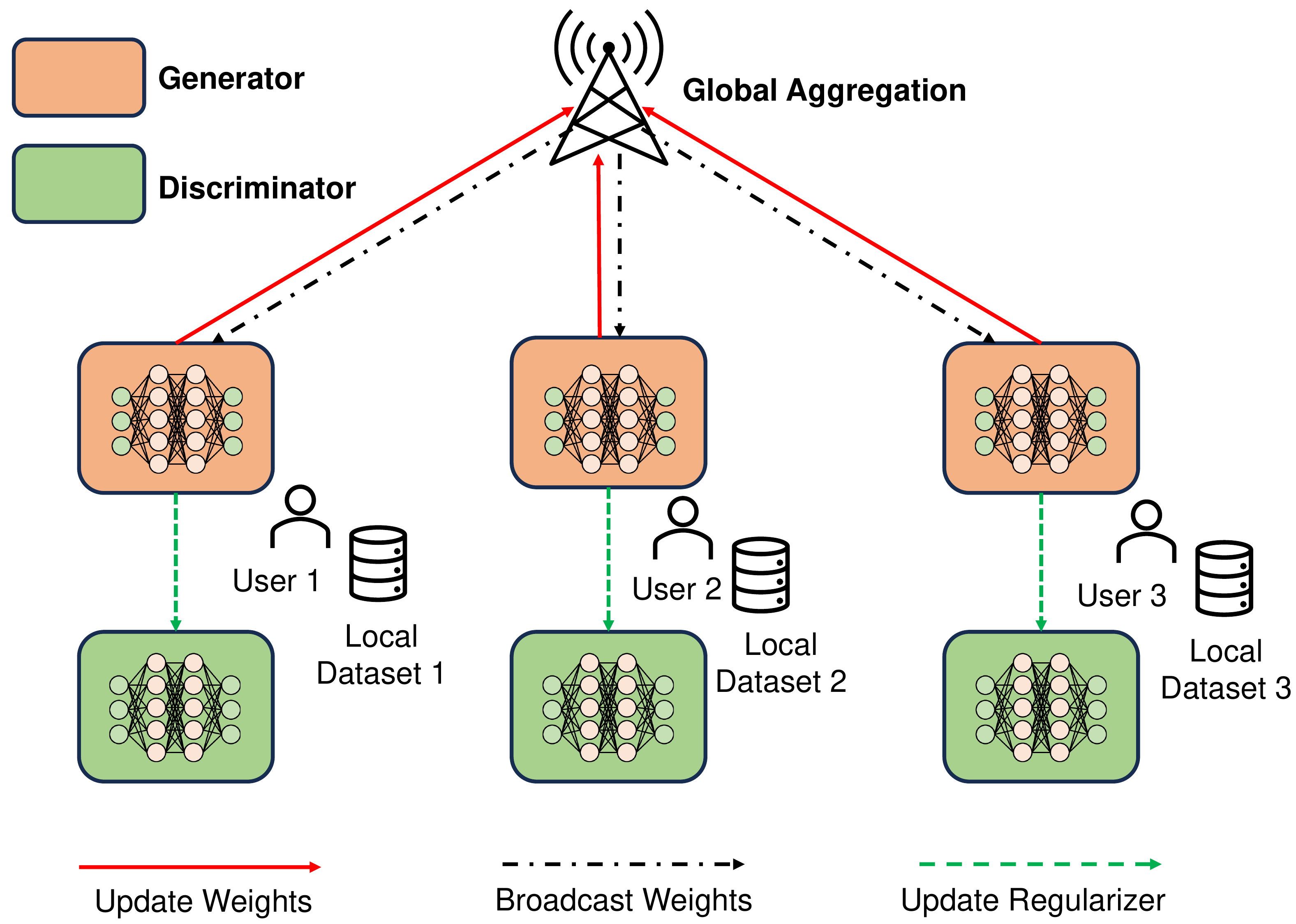}
	\caption{Architecture of proposed federated learning, in which only the generators' models are sent to the global server for aggregation. The discriminators then update their regularizer terms based on the global generator model's weights~\cite{guo2022federated}.}
	\label{Fig.federated_gan_channel_estimation}
\end{figure}

\begin{table*}[!]
	\centering
	\caption{Summary of GAI Approaches for Channel Equalization, Modeling, and Estimation}
	\begin{tabular}{|M{0.6cm}|m{1.2cm}|m{6cm}|m{7cm}|}
		\hline
		\textbf{No.} & \textbf{Problem} & \textbf{Drawbacks of TAI} & \textbf{Proposed Approach}\\
		\hline
		\hline
		\cite{caciularu2018blind} & \multirow{5}{*}{\parbox{1.5cm}{Channel Equalization}}  & Poor performance in blind channel equalization &  Use VAE to efficiently learn from unknown input impulse sequences.  \\ \cline{1-1}\cline{3-4}
		\cite{caciularu2020unsupervised} & & High complexity and not effective without using pilot symbols &   Use VAE to design a blind channel equalization that can model the unknown nonlinearity  \\ \cline{1-1}\cline{3-4}
		\cite{zou2023underwater} &  & Low performance under the scattering and absorption effects of underwater communications &  GAN is used to equalize the one-bit quantization's distortion as well as the negative effects of underwater channels.  \\ \cline{1-1}\cline{3-4}
		\cite{wu2023cddm} &  & Not effective in learning signal distributions &  Use diffusion models to eliminate channel noise  \\ \hline\hline
		\cite{o2019approximating} & \multirow{7}{*}{\parbox{1.5cm}{Channel Modeling}} & Only effective with simple channel models & Use GAN to learn the probability distribution functions of wireless channels, resulting in better channel response approximation  \\ \cline{1-1}\cline{3-4}
		\cite{wei2021variational} & & Suffer from the curse of dimensionality and can only be evaluated with a simple AWGN channel model &   Use VAE to learn the distribution of channel impulse responses and generate synthetic channel response samples with similar properties \\ \cline{1-1}\cline{3-4}
		\cite{zhang2021distributed} & & Limited by the lack of channel samples and environmental measurements &  Propose a distributed GAN architecture to allow UAVs to collaboratively approximate mmWave channel distributions \\ \cline{1-1}\cline{3-4}
		\cite{hu2022multi} & & Lack of training data and not effective in learning complex statistical relationships across different frequencies & Use GAN to generate random multi-cluster profiles that include all information of different frequencies  \\ \cline{1-1}\cline{3-4}
		\cite{sengupta2023generative} & & The collection of wireless channel data is costly and time-consuming & Propose a diffusion model based channel sampling approach to generate synthetic channel responses based on limited ground truth data  \\ \cline{1-1}\cline{3-4}
		\cite{rasheed2022lstm} & & Focus on estimating mmWave channel models for specific environments with limited applications &  Use GAN for mmWave channel modeling by effectively extracting useful CSI features in the spatial-temporal domain  \\ \hline\hline
		\cite{li2018generativees} & \multirow{9}{*}{\parbox{1.5cm}{Channel Estimation}} & High complexity and training overhead needed to obtain channel knowledge &  Use GAN to learn functions of channel covariance matrices and environment factors   \\ \cline{1-1}\cline{3-4}
		\cite{zhang2021generative} &  &  Cannot estimate channels in high-speed moving scenarios & Use GAN to learn and extract  channel time-varying features and then restore channel information\\ \cline{1-1}\cline{3-4}
		\cite{balevi2021wideband} & & Need to know or model the channel distribution & Use GAN to generate synthetic channel samples that have a similar distribution with a true but unknown channel  \\ \cline{1-1}\cline{3-4}
		\cite{hu2020channel} & & Poor performance and require a large number of simulated samples to train DNNs &  Use GAN to learn from receive signals and exploit Wasserstein distance to improve estimation accuracy without transmitting long pilot sequences.  \\ \cline{1-1}\cline{3-4}
		\cite{guo2022federated} & & Low privacy due to large CSI dataset exchanging & Each client uses the estimated CSI obtained by the least square estimator as the input data of GAN to approximate the channel's distribution  \\ \cline{1-1}\cline{3-4}
		\cite{zhang2021distributede} & & Do not focus on the characteristic of mmWave frequencies or A2G wireless links & Use GAN to learn the distribution of mmWave channels from multiple distributed datasets  \\ \cline{1-1}\cline{3-4}
		\cite{banerjee2022downlink} & & Require large datasets, long training time, and less effective under environmental variations  &  Develop a conditional GAN approach to generate channel covariance matrices for training \\ \cline{1-1}\cline{3-4}
		\cite{zhang2023channel} & & Do not adequately account for the dynamics and uncertainty of channels in large MIMO systems &  Use GAN to generate a more realistic channel image for more effective training under channel variations. \\ \hline\hline
	\end{tabular}
	\label{tab:Summary_channel_estimation_equalization}
\end{table*}

In addition, the authors in~\cite{zhang2021generative} reveal that conventional DL methods perform poorly in estimating fast time-varying and non-stationary channels. As such, they propose a GAN-based channel estimation approach that can accurately estimate wireless channels in high-speed railway systems. Specifically, the discriminator is responsible for learning and extracting time-varying features of railway communication channels while the generator aims to determine the training data's implicit mapping function. Through simulations, the authors show that the probability density curve of the estimated channel data is highly similar to that of the ground truth channel responses, indicating the effectiveness of GAN in estimating wireless channels.

The aforementioned GAN-based solutions and many others in the literature are designed in a centralized learning manner which may not be feasible in large-scale scenarios. To tackle this practical challenge, the authors in~\cite{guo2022federated} propose a federated GAN solution for channel estimation in a distributed manner, as illustrated in Fig.~\ref{Fig.federated_gan_channel_estimation}. In particular, each client uses the estimated CSI obtained by the least square estimator as the input data of GAN to learn the distribution of channels. After that, the generator parameters are transmitted to the server for aggregation. To improve federated learning performance, each client's discriminator will be dynamically adjusted by using regularizers based on the global generator's weights. Extensive simulations suggest that the proposed federated GAN approach is superior to conventional estimators as well as state-of-the-art DL-based channel estimation. For example, at SNR=5 dB, GAN can achieve a normalized mean-squared error (NMSE) of $10^{-2}$ while the ChannelNet proposed in~\cite{soltani2019deep} can only obtain an NMSE of around 0.5. To further reduce the communication overhead, model compression and multiple tasks design can be considered to make the proposed federated GAN approach more effective.

%\cite{caciularu2018blind, li2018generativees, o2019approximating, caciularu2020unsupervised, zhang2021generative, wei2021variational, balevi2021wideband, zhang2021distributed, hu2020channel, guo2022federated, hu2022multi, ye2022channel, zhang2021distributede, sengupta2023generative, rasheed2022lstm, banerjee2022downlink, zou2023underwater, zhang2023channel, wu2023cddm}

\subsection{Physical Layer Security}

Physical layer security (PLS) is another important research area in wireless communication systems. In general, PLS refers to techniques that enhance the security of wireless communications at the physical layer by leveraging the inherent randomness of wireless communication channels. Major problems in PLS include anti-jamming, anti-eavesdropping, signal authentication, and device identification. With recent advancements in DNNs, DL has been widely adopted to improve the security at the physical layer of wireless communication systems. However, conventional DL-based approaches face various challenges due to the dynamics and uncertainty of diverse physical layer attacks. Specifically, the DL model is usually trained on the dataset of a specific environment, and thus it cannot work well in new conditions. This is an essential problem as attackers can always change their attack strategies to maximize the disruption. In addition, conventional DL-based approaches require large datasets to obtain good detection performance. However, it is difficult to collect sufficient labeled data from physical layer attacks due to their randomness and dynamics~\cite{cai2020spectrum, tang2020jamming, han2021better}. More importantly, conventional DL models are vulnerable to adversarial attacks~\cite{merchant2019securing, de2022multi}. Minor perturbations in the input data can fool DNNs and consequently make conventional DL-based approaches less effective in dealing with physical layer attacks. Finally, conventional DL models perform poorly with time-varying channels in low SNR regions and when prior information about attackers is not readily available~\cite{yang2022simple, meng2022physical}.

\begin{figure}[!]
	\centering
	\includegraphics[scale=0.23]{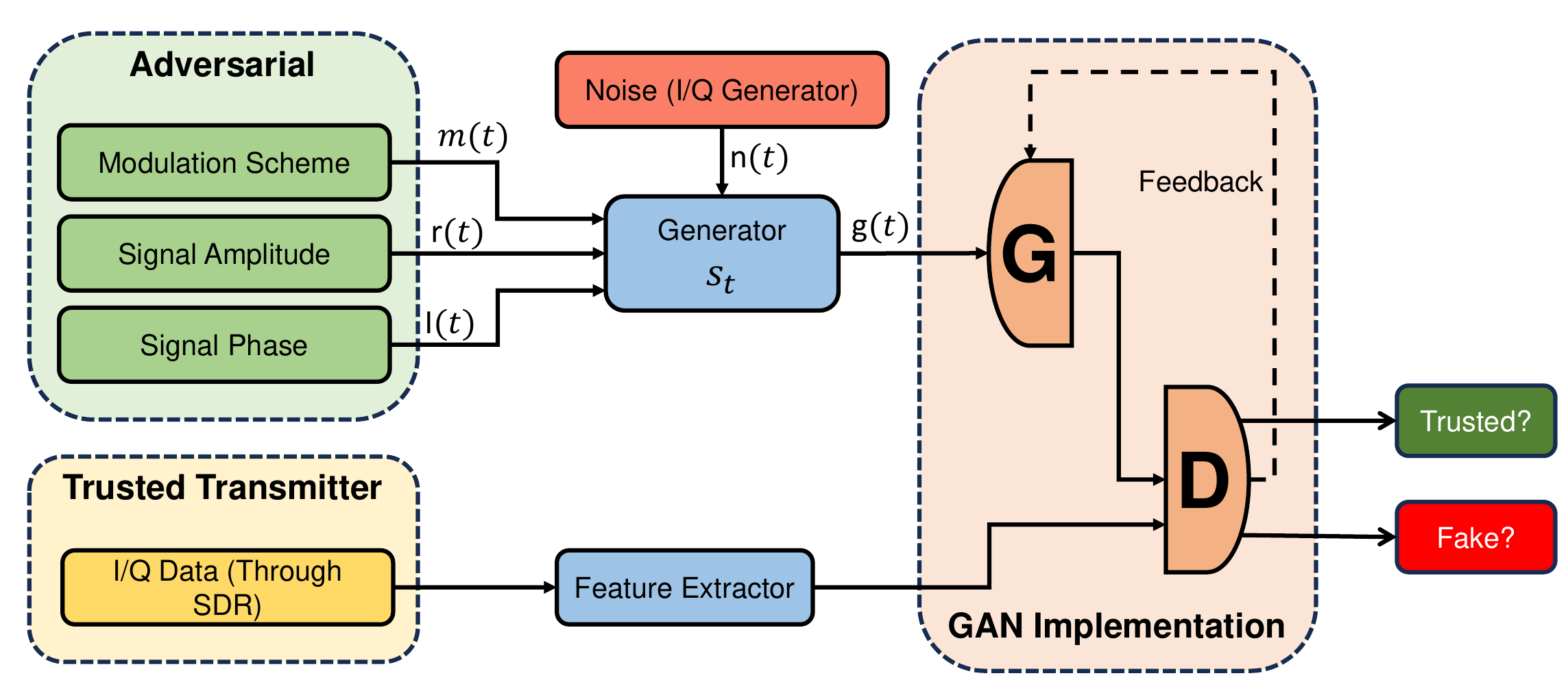}
	\caption{Proposed GAN architecture for authenticating RF transmitters in which the generator uses RF signals generated from adversaries as its input to generate synthetic data $g(t)$ and the discriminator takes input from both the generator and ``trusted'' transmitters to learn the differences between real and fake RF signals~\cite{roy2019detection}.}
	\label{Fig.gan_authenticate_RF_transmitters}
\end{figure}

These challenges in physical layer security can be efficiently addressed by GAI, as summarized in Table~\ref{tab:physical_layer_security}. As discussed, GAI can be used for anomaly detection by generating data that is defined to be normal and then flagging input data that deviates significantly from these definitions. In addition, GAI has been demonstrated to be effective in uncertainty estimation as well as domain adaptation~\cite{liu2023deep} which are critical capabilities to deal with physical layer security threats. For example, a GAN-based solution is proposed in~\cite{toma2020ai} for abnormality detection at the physical layer in cognitive radio networks. In particular, the proposed GAN approach is used to generalize the state vectors extracted from spectrum representation data to learn the dynamic behavior of wideband signals. Based on these state vectors, abnormal signals can be distinguished from legitimate signals. Similarly, the authors in~\cite{han2021primary} and~\cite{han2021better} aim to prevent jamming attacks as well as interference from secondary users in cognitive radio networks. They first highlight that conventional DL-based anti-jamming approaches give poor performance when spectrum data is not sufficient. Unfortunately, collecting and labeling spectrum data in the presence of jamming attacks are time-consuming and costly. To address this practical issue, the authors propose to use GAN to generate synthetic spectrum data that can help a DRL algorithm to effectively learn and obtain the optimal dynamic spectrum anti-jamming access policy. Extensive simulations then demonstrate that the proposed GAN can help to avoid complex jamming attacks and outperform conventional DRL-based approaches with incomplete spectrum information. The lack of training data problem of conventional physical layer security approaches is also discussed and addressed by using GAN in~\cite{tang2020jamming},~\cite{han2020radio},~\cite{yang2022simple}, and~\cite{yang2022generative}.

Differently, the work in~\cite{roy2019detection} aims to authenticate radio frequency (RF) transmitters by using GAN. The authors first highlight that conventional ML techniques cannot be straightforwardly applied to RF systems due to the dynamics and uncertainty of RF signals. More importantly, these ML techniques may perform poorly in the presence of intelligent adversaries that can spoof transmitters and inject interference into the target channels, making it more challenging to capture the unique properties of the transmitters. For that, the authors propose a GAN-based approach to efficiently authenticate RF transmitters as GAN is well known for its capability in dealing with adversarial situations, as shown in Fig.~\ref{Fig.gan_authenticate_RF_transmitters}. In particular, the GAN's generator will use RF signals generated from adversaries as its input to generate synthetic data $g(t)$. On the other hand, the discriminator learns from signals of both ``trusted'' transmitters and the generator to identify the differences between real and fake RF signals. In this way, the proposed solution can achieve a detection accuracy of 99\% which is much higher than those of CNN and DNN approaches, i.e., 81.6\% and 96.6\%, respectively. Similarly, the authors in~\cite{zhou2021radio} also point out that the time-varying characteristics of wireless channels introduce more difficulties to conventional DL-based approaches in detecting abnormal RF signals. In contrast, GAN, with its capabilities of anomaly detection and uncertainty estimation, can deal with this issue effectively.

In~\cite{li2021gnss}, the authors study a more challenging scenario where spoofing signals are identical or similar to real signals. Specifically, spoofing attacks in the global navigation satellite system are considered in which spoofing signals are similar to legitimate satellite signals in terms of pseudo-code phase and carrier Doppler values but have much stronger power to lure the receiver to track them instead of real signals. Consequently, existing detection methods, including DL-based solutions, cannot effectively distinguish between legitimate satellite signals and spoofing signals. The authors then design a GAN network that is trained on a large dataset of authentic satellite signals to accurately learn their distributions. Simulation results indicate that by using GAN, the authors can obtain better detection performance than using the conventional CNN network. Differently, the authors in~\cite{han2022novel} leverage GAN as an effective tool for physical layer key generation. It is well known that wireless communications are susceptible to radio attacks such as eavesdropping and tampering due to their broadcast nature. Meanwhile, conventional cryptography techniques in the upper layers may not be feasible for wireless devices due to their computational complexity, especially in IoT networks. By leveraging the inherent uncertainties of the physical communication channels, physical layer key generation has been widely adopted. In general, DL is superior to conventional approaches in extracting symmetric keys from reciprocal channel responses. However, the authors in~\cite{han2022novel} reveal that conventional DNNs are unpredictable for physical layer key generation. In addition, it is challenging to apply the extracted high-dimensional features to generate the physical layer key. Therefore, they propose a new key generation method based on GAN that can efficiently extract features of legitimate nodes.

\begin{figure}[!]
	\centering
	\includegraphics[scale=0.27]{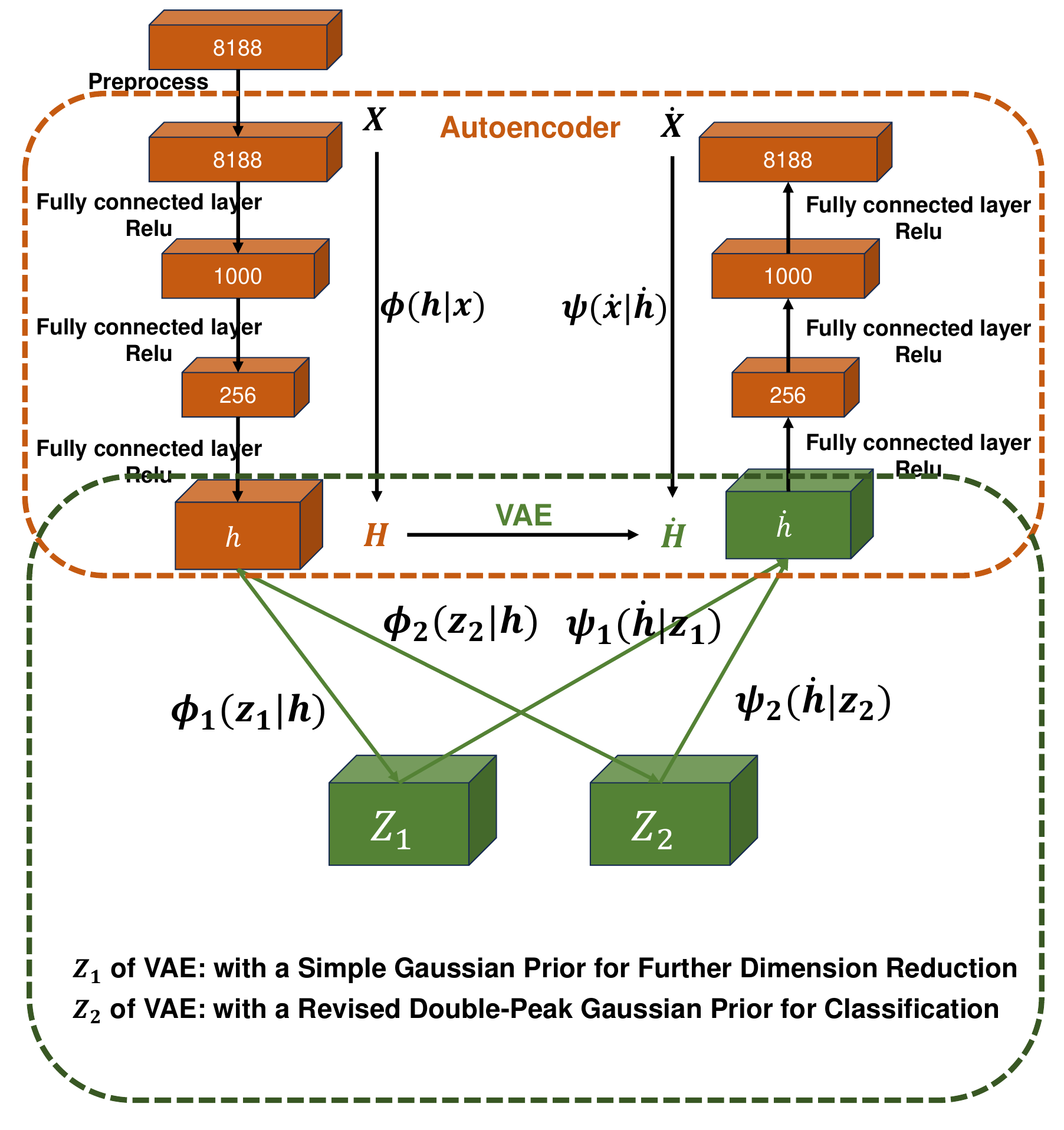}
	\caption{Proposed GAN architecture for authenticating RF transmitters in which the VAE is used as a high-level classifier with two hidden units $Z_1$ and $Z_2$. $Z_1$ is constructed based on encoder $\phi_1$ and decoder $\psi_1$ with a Simple Gaussian Prior for dimension reduction and channel impulse response reproduction. $Z_2$ is constructed based on encoder $\phi_2$ and decoder $\psi_2$ with a revised double-peak Gaussian Prior for authentication~\cite{roy2019detection}.}
	\label{Fig.vae_authentication}
\end{figure}

\begin{table*}[!]
	\centering
	\caption{Summary of GAI Approaches for Physical Layer Security}
	\begin{tabular}{|M{0.6cm}|m{1cm}|m{6cm}|m{7.1cm}|}
		\hline
		\textbf{No.} & \textbf{Problem} & \textbf{Drawbacks of TAI} & \textbf{Proposed Approach}\\
		\hline
		\hline
		\cite{gong2019generative} & \multirow{27}{*}{\parbox{1.5cm}{Physical Layer Security}}  & Inaccurate representations for the individual emitters &  Use GAN to extract hidden information in original signals to improve identification performance     \\ \cline{1-1}\cline{3-4}
		\cite{shi2019generative} & & Cannot generate high-quality synthetic spoofing signals  & Use GAN to generate spoofing signals that are similar to legitimate signals  \\ \cline{1-1}\cline{3-4}
		\cite{merchant2019securing} & & TAI-based approach can be cracked by using GAN &  Use GAN to augment the training dataset of the classifier with adversarial samples generated from another GAN network \\ \cline{1-1}\cline{3-4}
		\cite{erpek2018deep} & & Cannot generate high-quality synthetic samples & Use GAN to help jammers to generate training data to improve attack performance \\ \cline{1-1}\cline{3-4}
		\cite{roy2019detection} & & Not effective when detecting rogue RF transmitters and classifying trusted ones & Use a generative model to generate fake signals that are trained together with real signals by a discriminative model to better identify trusted ones  \\ \cline{1-1}\cline{3-4}
		\cite{roy2019generative} & & Not effective in learning data distributions & Use GAN to provide accurate estimation of complex probability distributions from smallish sized datasets  \\ \cline{1-1}\cline{3-4}
		\cite{fan2020deceptive} & &Cannot generate deceptive jamming templates under constraints & Use GAN to adaptively generate refined deceptive jamming templates based on various factors such as azimuth angles, angles, and target types. This can help to protect a specific area from observation and detection by adversarial radars  \\ \cline{1-1}\cline{3-4}
		\cite{toma2020ai} & & Only effective with a specific type of low dimensional data & Use GAN to effectively learn from high dimensional data of spectrum representation samples \\ \cline{1-1}\cline{3-4}
		\cite{cai2020spectrum} & & Lack of spectrum data & Use GAN to generate incomplete spectrum data in multiple jamming patterns  \\ \cline{1-1}\cline{3-4}
		\cite{germain2020physical} & & Not effective in learning distributions of received signals & Use GAN to learn the distribution of received channel data to authenticate a transmitting device  \\ \cline{1-1}\cline{3-4}
		\cite{tang2020jamming} & & Lack of training data & Use GAN to generate more samples and use VAEs to learn the latent space of continuous signal samples.  \\ \cline{1-1}\cline{3-4}
		\cite{han2020radio} & &Lack of training data about malicious transmitters &  Use GAN to learn the data of trusted transmitters to extract RF fingerprint \\ \cline{1-1}\cline{3-4}
		\cite{han2021better} & &Lack of spectrum data  & Use GAN to generate missing spectrum data  \\ \cline{1-1}\cline{3-4}
		\cite{germain2021mobile} & & Not effective in dealing with the dynamics of wireless channels &  Use GAN and LSTM to learn and predict CSI elements' magnitude \\ \cline{1-1}\cline{3-4}
		\cite{zhou2021radio} & &Time-varying characteristics of wireless channels make the prediction unreliable  & Incorporating an encoder network into the original GAN to reconstruct the spectrogram  \\ \cline{1-1}\cline{3-4}
		\cite{li2021gnss} & & Most detection methods cannot effectively detect spoofing jamming if spoofing signals are similar to authentic signals & Design a GAN network that is trained on a large dataset of authentic satellite signals to accurately learn their distribution  \\ \cline{1-1}\cline{3-4}
		\cite{shi2020generative} & &Cannot effectively use to perform attacks  & Use GAN to construct synthetic RF signals that are similar to legitimate signals  \\ \cline{1-1}\cline{3-4}
		\cite{han2022novel} & &Difficult to apply to the key generation in the physical layer  & Propose a key generation method based on GAN to extract features efficiently between legitimate nodes  \\ \cline{1-1}\cline{3-4}
		\cite{barnes2022scalable} & & Not effective in anomaly detection as GAI  &  Use GAN to identify unrecognized patterns on the model outputs and associated sequenced metadata \\ \cline{1-1}\cline{3-4}
		\cite{han2021primary} & &Poor performance when spectrum data is not sufficient  & Use GAN to generate synthetic spectrum data that can help DRL to effectively learn and obtain the optimal dynamic spectrum anti-jamming access policy  \\ \cline{1-1}\cline{3-4}
		\cite{yang2022generative} & &Lack of labeled data  & Use GAN to learn the distribution of collected signals \\ \cline{1-1}\cline{3-4}
		\cite{meng2022physical} &  &Need attackers' information for training & Use VAEs to extract valuable features of high-dimensional channel impulse responses for authentication  \\ \hline\hline
	\end{tabular}
	\label{tab:physical_layer_security}
\end{table*}

Besides GAN, VAEs can also be adopted for physical layer security. For instance, the authors in~\cite{meng2022physical} propose a hierarchical VAE-based approach for physical layer authentication in complex scenarios such as industrial IoT systems. The authors state that ML has been widely adopted for physical layer authentication to analyze and extract complicated properties of wireless channels for authenticating wireless devices. Nevertheless, these methods usually require information about attackers available in advance to obtain good detection performance which is not the case in practice. As such, the authors develop a new hierarchical VAE architecture based on autoencoder and VAEs for efficient physical layer authentication with no prior channel information of attackers, as illustrated in Fig.~\ref{Fig.vae_authentication}. In particular, the VAE is used as a classifier, consisting of two hidden units $Z_1$ and $Z_2$. $Z_1$ is constructed based on encoder $\phi_1$ and decoder $\psi_1$ with a Simple Gaussian Prior for dimension reduction and channel impulse response reproduction. On the other hand, $Z_2$ is constructed based on encoder $\phi_2$ and decoder $\psi_2$ with a revised double-peak Gaussian Prior for authentication. The conventional autoencoder is used to further reduce the dimension of input data. Finally, a new loss function is designed for the VAE module considering both the Simple Gaussian Prior and the double-peak Gaussian Prior distributions for further security and robustness enhancement. In this way, the proposed solution can efficiently capture important features of high-dimensional channel impulse responses for better authentication performance. The authors then show that the proposed solution can improve the authentication performance by 17.18\% compared to a conventional ML approach in~\cite{xia2021multiple}.

Due to the ability of generating synthetic data that is similar to real data, GAI can also be used by adversaries to perform different types of physical layer attacks~\cite{erpek2018deep, merchant2019securing, shi2020generative, de2022multi}. For example, the authors in~\cite{shi2020generative} use a GAN network to generate synthetic wireless signals that cannot be distinguished from legitimate signals by conventional approaches. Experiments then show that by using GAN the authors can improve the attack performance. Similarly, the authors in~\cite{de2022multi} recruit GAN to perform adversarial attacks. In particular, GAN is used to generate crafted imperceptible perturbations to cause wrong classifications of a DL-based modulation recognition approach. Through extensive simulations, the authors then indicate that the proposed GAN-based adversarial attack can reduce the accuracy of the DL-based modulation classifier more than jamming and other adversarial attacks. For instance, at 0 dB perturbation-to-noise ratio, the proposed techniques can reduce the detection performance by 37\% at SNR=10 dB, by 56\% at SNR=0 dB, and by 7\% at SNR=-10 dB. In addition, the authors in~\cite{erpek2018deep} demonstrate that GAN can help a jammer to effectively jam a target wireless channel by generating more training data to help the jammer better learn the defense policy of the legitimate receiver. To deal with these GAN-based attacks, the authors in~\cite{merchant2019securing} propose to use another GAN network to augment the training dataset of the classifier with adversarial samples generated from adversaries' GAN networks. Simulation results then show that by augmenting the training data with GAN the authors can effectively improve the classification accuracy under GAN-based adversarial attacks.

%\cite{gong2019generative, shi2019generative, merchant2019securing, erpek2018deep, roy2019detection, roy2019generative, fan2020deceptive, toma2020ai, cai2020spectrum, germain2020physical, tang2020jamming, han2020radio, han2021better, germain2021mobile, zhou2021radio, li2021gnss, shi2020generative, xu2022waveform, han2022novel, barnes2022scalable, han2022anti, de2022multi, xu2022colluding, han2021primary, yang2022simple, yang2022generative, meng2022physical}

\subsection{Intelligent Reflecting Surface (IRS)}

Recently, IRS has been emerging as a promising technology to significantly improve energy efficiency and spectrum utilization with low-cost and low-power hardware~\cite{wei2021model, wei2022channel}. In particular, a typical IRS consists of a large number of reconfigurable metasurface elements that can be adjusted to control the amplitude responses and phase shifts of the incoming incident electromagnetic waves. By coordinating the phase shifts and amplitudes across the array of elements, an IRS can reconfigure wireless channels and obtain high beamforming gain in a desired direction, creating a favorable wireless signal propagation environment. However, accurate CSI information and underlying channel models must be obtained to leverage these advantages of IRS~\cite{ye2022channel}. Unfortunately, it is challenging to acquire BS-IRS and IRS-UE channels separately without the help of RF chains. In addition, the cascaded channel of BS-IRS and IRS-UE links is very high-dimensional due to the high number of reflecting elements. To overcome these issues, various DL-based channel estimation and channel modeling approaches have been proposed in the literature. Nevertheless, these approaches cannot accurately estimate IRS channels since they use a general loss function that is not well designed for IRS, leading to poor estimation performance~\cite{ye2022channel}. In addition, conventional DL-based approaches can only learn a limited number of channel parameters and one-dimensional channel impulse responses.

\begin{table*}[!]
	\centering
	\caption{Summary of GAI Approaches for Intelligent Reflecting Surface (IRS)}
	\begin{tabular}{|M{0.6cm}|m{1.8cm}|m{5cm}|m{7.3cm}|}
		\hline
		\textbf{No.} & \textbf{Problem} & \textbf{Drawbacks of TAI} & \textbf{Proposed Approach}\\
		\hline
		\hline
		\cite{wei2021model, wei2022channel} & Channel Modeling  & Require in-depth domain knowledge and lack of training data &   Use GAN to generate high-dimensional channel samples    \\ \cline{1-4}
		\cite{jin2021multiple} & \multirow{3}{*}{\parbox{1.5cm}{Channel Estimation}}  & Lack of observational dimensions and modeling capabilities  & Use GAN to remove noise from the estimated channel matrix      \\ \cline{1-1}\cline{3-4}
		\cite{li2023uplink} &   & Not effective with high channel dimensions & Use GAN to learn the channel distribution with LS estimation as conditional input      \\ \cline{1-1}\cline{3-4}
		\cite{ye2022channel} &   & Use a general loss function that is difficult to make the estimated IRS channels more accurate &  Use GAN to approximate cascaded channels by taking received signals as conditional information    \\ \cline{1-4}
		\cite{naeem2023joint} & IRS Deployment Design & Not effective in dealing with the dynamics of 6G networks & Use GAN to support DRL by learning the action-value that is near to target-action values, resulting in a more stable learning process  \\ \hline\hline
	\end{tabular}
	\label{tab:IRS}
\end{table*}

To tackle the above issues of conventional DL-based approaches, GAI has been adopted in various studies, as summarized in Table~\ref{tab:IRS}. For example, the authors in~\cite{wei2021model} and~\cite{wei2022channel} develop a model-driven framework based on GAN for channel modeling in IRS-aided wireless communication systems. To make GAN learn the channel distribution more effectively, the authors incorporate the structure of the cascaded BS-IRS and IRS-UE channels into the generator of the proposed GAN architecture. More specifically, the generative model now has three nodes: (i) BS-IRS node to learn BS-IRS channel distribution, (ii) IRS-UE node to learn IRS-UE channel distribution, and (iii) cascading node to combine the outputs. The discriminative model is then used to distinguish between the generated channel samples and the real BS-IRS-UE channel samples. Moreover, the authors adopt Wasserstein distance~\cite{arjovsky2017wasserstein} to design a new loss function for the proposed GAN model for more stable training. In this way, the proposed solution can achieve much better performance than existing solutions using CNNs and fully-connected neural networks (FNNs), as demonstrated in the simulation results. Differently, the authors in~\cite{ye2022channel} use a conditional GAN architecture for channel estimation in IRS-aid wireless communications. In particular, the proposed GAN takes the received signals as its conditional information to generate channel responses with certain characteristics. Then, the discriminator and the generator compete with each other to obtain an adaptive loss function, making the generated channels similar to the original channels. With its capability in learning data distribution effectively, the proposed GAN architecture can achieve much better channel estimation performance compared to conventional DL-based methods as demonstrated in extensive simulations. For instance, at 5 dB SNR, the NMSE of GAN is around ten times less than that of the ChannelNet architecture proposed in~\cite{elbir2020deep}. Applications of GAN for channel estimation in IRS-aided wireless communications are also studied in~\cite{jin2021multiple} and~\cite{li2023uplink} where GAN-based convolutional blind denoising and conditional GAN are adopted to obtain accurate CSI for IRS-aided systems, respectively.

GAN can also be used for the deployment design and phase shift optimization of IRS. For instance, the authors in~\cite{naeem2023joint} aim to jointly optimize the placement and reflecting beamforming matrix of an IRS-assisted 6G network. The authors first develop a deep reinforcement learning (DRL) framework to interact with the system and gradually learn an optimal joint policy. However, due to the reward function's randomness, the proposed DRL framework cannot learn all the dynamics and uncertainty of the considered IRS system effectively. To overcome this issue, the authors propose to use GAN to identify the action-value that is close to target-action values, resulting in a more stable learning process. Specifically, the generator aims to generate actions (e.g., adjusting phase shift, coordinates, and beamform) for the DRL agent that are mapped to the original dataset. Then, these generated experiences and the original dataset are stored in a relay buffer. After that, the discriminator randomly takes a number of samples in the relay buffer as its input to learn how to distinguish the generated experiences from the generator and real samples from the original datasets. Simulation results then demonstrate that the proposed GAN architecture can help to improve the accuracy of DRL by 45\%. To allow multiple IRSs to work collaboratively, the proposed approach can be extended by considering a multiagent GAN-based DRL framework.

\subsection{Beamforming}
\label{subsec:beamforming}

\begin{figure*}[!]
	\centering
	\includegraphics[scale=0.18]{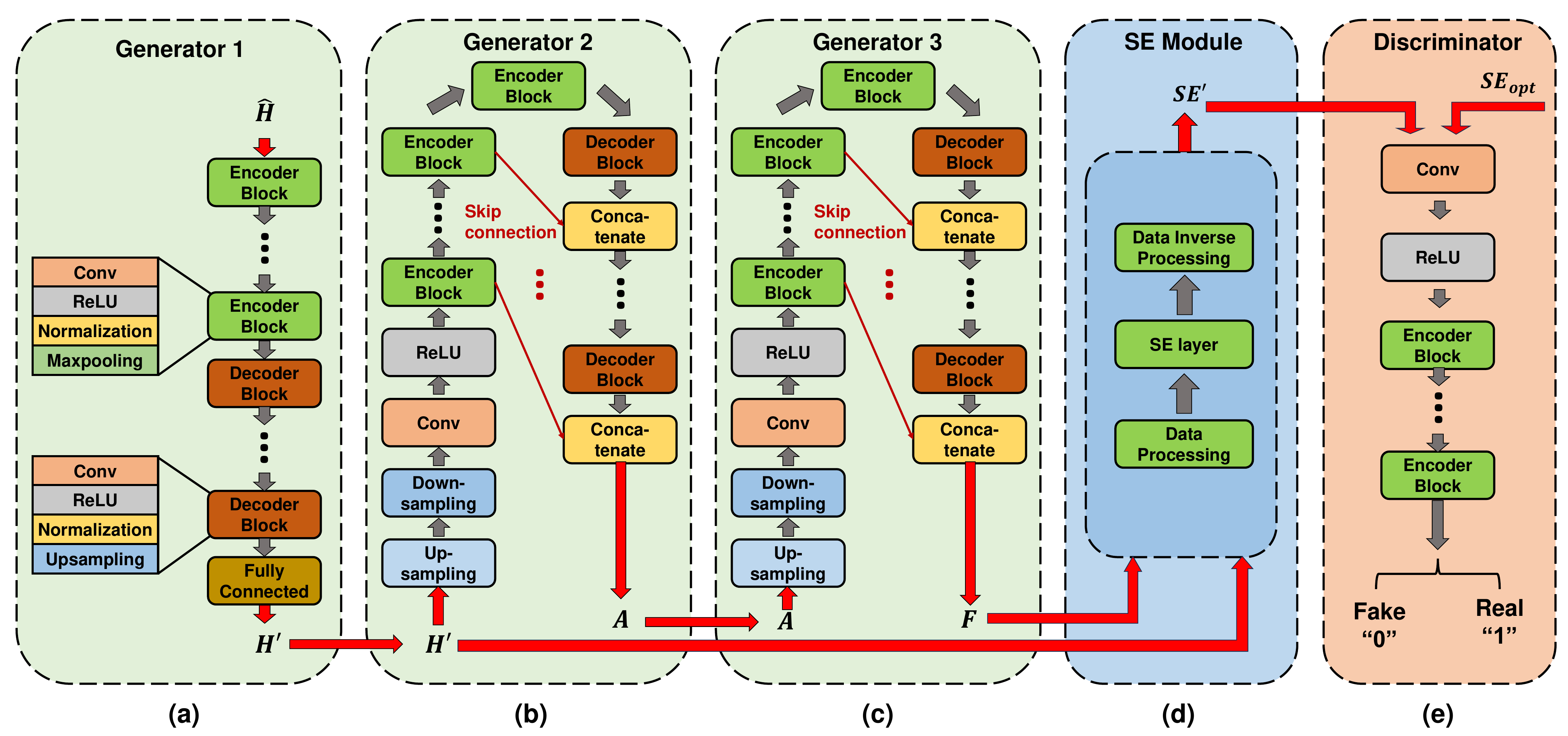}
	\caption{Multi-GAN architecture for beamforming with (a) Generator G1, (b) Generator G2, (c) Generator G3, (d) the spectrum efficiency module, and (e) the discriminator~\cite{pang2022mggan}.}
	\label{Fig.gan_for_beamforming}
\end{figure*}

\begin{table*}[!]
	\centering
	\caption{Summary of GAI Approaches for Beamforming}
	\begin{tabular}{|M{0.6cm}|m{1.8cm}|m{4cm}|m{8.4cm}|}
		\hline
		\textbf{No.} & \textbf{Problem} & \textbf{Drawbacks of TAI} & \textbf{Proposed Approach}\\
		\hline
		\hline
		\cite{ngo2020deep} & \multirow{5}{*}{\parbox{1.5cm}{Beamforming}}  & Lack of training data &  Use GAN to generate additional data for beam prediction    \\ \cline{1-1}\cline{3-4}
		\cite{hussain2021adaptive}, \cite{hussain2021learning} &   & Not effective in learning data distribution  &  Use VAE to approximate the probabilistic model of beam dynamics    \\ \cline{1-1}\cline{3-4}
		\cite{balevi2021unfolded} &   & High feedback overhead and high complexity  & Use GAN to reconstruct a low-dimensional channel fed back from the receiver to perform hybrid beamforming at the transmitter    \\ \cline{1-1}\cline{3-4}
		\cite{pang2022mggan} &   & Suffer from the rank-deficient problem  & Use a GAN architecture with three generators to recover (i) rank-deficient channels, obtain (ii) analog beamforming and (iii) digital beamforming matrices.       \\ \hline\hline
	\end{tabular}
	\label{tab:beamforming}
\end{table*}

In wireless communications, beamforming is a key technology to improve signal quality and transmission coverage. To do that, the transmitter, e.g., BSs, employs an array of individual antenna elements in which each element can adjust the phase of signals to make signals at specific angles experience destructive interference while others encounter constructive interference. However, it is challenging to obtain optimal beamforming policies due to the high computational complexity and excessive feedback overhead, especially in systems with large antenna arrays like mmWave and massive MIMO communication systems~\cite{hussain2021adaptive, balevi2021unfolded}. DL can be used to tackle this problem but it requires a large amount of training data and cannot efficiently deal with the dynamics and uncertainty of wireless communications. Several researchers have been adopting GAI as an alternative approach and achieving promising results, as summarized in Table~\ref{tab:beamforming}. For example, the authors in~\cite{balevi2021unfolded} propose to use GAN to reconstruct low-dimensional channel feedback from the receiver to perform hybrid beamforming at the transmitter, resulting in low communication overhead. Specifically, the generator of the proposed GAN architecture is first pretrained offline with channel samples generated by a geometric channel model to learn the channel structure and correlations. In the online phase, the receiver tries to compress the channel matrix to a low-dimensional vector and feeds it back to the proposed GAN architecture at the transmitter to recover the channel matrix that will be used for beamforming design. The proposed GAN solution can help to reduce 2,048 complex channel elements to just 15 real values while maintaining good communication performance as demonstrated in simulations. One possible extension of this work is to extend the considered model/method to extremely-large massive MIMO or holographic MIMO.

Differently, the authors in~\cite{pang2022mggan} consider the beamforming design in massive MIMO systems with large antenna arrays where only rank-deficient CSI can be obtained. This rank-deficient problem has not been fully solved in the literature using conventional techniques. For that, the authors propose a multi-GAN architecture for hybrid beamforming design under rank-deficient channels. Specifically, the authors employ three generators in the proposed GAN architecture in which generator G1 is used to recover rank-deficient channels, generator G2 is used for analog beamforming, and generator G3 is used for hybrid beamforming, as illustrated in Fig.~\ref{Fig.gan_for_beamforming}. Generator G2 takes the estimated rank-deficient channel $\mathbf{H}'$ as its input to generate analog beamforming $\mathbf{A}$ which is the input of generator G3. Generator G3 then estimates hybrid beamforming $\mathbf{F}$. After that, $\mathbf{H}'$ and $\mathbf{F}$ are fed into the spectrum efficiency module to calculate the average spectrum efficiency $\mathbf{SE}'$. The generator then learns from $\mathbf{SE}'$ and the real spectrum efficiency to improve the training processes of generators G1 and G3. Extensive simulations then demonstrate that the proposed multi-GAN architecture can improve the beamforming performance by 47.49\% compared to a conventional CNN-based method.

\subsection{Joint Source Channel Coding (JSCC)}
\label{subsec:JSCC}

\begin{table*}[!]
	\centering
	\caption{Summary of GAI Approaches for Joint Source Channel Coding (JSCC)}
	\begin{tabular}{|M{0.6cm}|m{1cm}|m{5cm}|m{8.2cm}|}
		\hline
		\textbf{No.} & \textbf{Problem} & \textbf{Drawbacks of TAI} & \textbf{Proposed Approach}\\
		\hline
		\hline
		\cite{saidutta2021joint} & \multirow{5}{*}{\parbox{1.5cm}{JSCC}}  & Not effective in dealing with the complexity and discontinuity of source data distributions & VAE's encoder converts the source data into a low-dimensional latent space while the VAE's decoder tries to recover it to original data for JSCC     \\ \cline{1-1}\cline{3-4}
		\cite{dai2022nonlinear} &   & Not effective when source dimension increases &  Use VAE to learn the source distribution by considering the noise channel as a sample of latent variables    \\ \cline{1-1}\cline{3-4}
		\cite{erdemir2023generative} &   & Have significant losses of perceptual quality for the edge cases &  Use two GAN-based networks to recover the distorted reconstructions of a DL-based JSCC and to produce the latent and noise inputs for the StyleGAN-2, respectively   \\ \cline{1-1}\cline{3-4}
		\cite{li2023joint} &   & Not stable for multivariate Gaussian source over Gaussian multiple access channels & Propose a VAE-based JSCC system with added distribution restrictions on the loss function to avoid falling into the local minimum in specific regions     \\ \cline{1-1}\cline{3-4}
		\cite{niu2023hybrid} &   & Suffer from the cliff effect & Use diffusion models as a generative refinement component to enhance the reconstruction's perceptual quality   \\ \cline{1-1}\cline{3-4}
		\cite{ye2023lightweight} &   & High complexity & Use a GAN compression method based on intermediate feature distillation       \\ \hline\hline
	\end{tabular}
	\label{tab:JSCC}
\end{table*}

Coding plays a crucial role in wireless communications to mitigate the negative effects of channel noise, interference, and fading. Traditionally, the transmitter performs source coding for compression and channel coding for error correction, separately, making it difficult to optimize the spectrum usage. By combining the functions of source coding and channel coding into a single process, JSCC can leverage the statistical characteristics of the source and the channel to design a more efficient coding method. JSCC also helps to reduce the overall complexity of wireless communication systems since it requires one encoder and one decoder only. However, the complexity and discontinuity of source data distributions introduce challenges to the design of JSCC. For that, the authors in~\cite{saidutta2021joint} propose a novel JSCC approach based on VAEs over additive noise analog channels. Specifically, the proposed VAE's encoder is used to convert source data into a low-dimensional latent space while the VAE's decoder recovers it to original data for JSCC. More importantly, the authors study that when the channel dimension is smaller than the source dimension, the encoding of two neighboring source samples needs to be near each other for good encoding performance. Therefore, multiple encoders are employed, and one of them will be selected for sample encoding on a specific side of the discontinuity. Experiments then demonstrate that using the proposed VAE-based JSCC method can help to increase the average peak SNR (PSNR) by nearly 3 dB compared to conventional CNN-based approaches.

Recently, JSCC has been emerging as an effective technology for semantic communications. However, in~\cite{dai2022nonlinear}, the authors highlight that when the source dimension increases, e.g., large-scale images, the performance of DL-based JSCC methods degrades significantly. Moreover, when the channel bandwidth ratio increases, these methods provide poor coding gain as they cannot learn the source distribution to determine patch-wise variable-length transmissions. To tackle these issues, the authors design a JSCC architecture based on VAEs in which the noise channel is viewed as a sample of latent variables. In this way, the proposed architecture can effectively learn the source distribution to provide a more effective coding mechanism. Experiments then show that the proposed solution can achieve up to 28.91\% bandwidth saving or a PSNR gain of 2.64 dB on the CIFAR10 dataset while the conventional deep JSCC increases the bandwidth cost by up to 54.31\%. Similarly, the authors in~\cite{erdemir2023generative} also consider JSCC for semantic image transmissions. The authors study that DL-based JSCC possesses significant perceptual quality losses in edge scenarios. Therefore, they propose two novel JSCC schemes based on GAN, namely InverseJSCC and GenerativeJSCC. InverseJSCC aims to recover the distorted reconstructions of a DL-based JSCC model via solving an inverse optimization problem using a pre-trained style-based GAN architecture. In contrast, in GenerativeJSCC, GAN is used as the decoder to produce latent and noise inputs for the StyleGAN-2~\cite{karras2020analyzing} generator. By jointly training the encoder and GAN decoder, GenerativeJSCC can outperform DL-based JSCC methods in terms of perceptual quality and distortion, as demonstrated by extensive simulations. GAI has also been adopted in other studies as summarized in Table~\ref{tab:JSCC}.

\begin{table*}[!]
	\centering
	\caption{Summary of GAI Approaches for CSI feedback, radio map estimation, and channel delay estimation}
	\begin{tabular}{|M{0.6cm}|m{1.8cm}|m{4cm}|m{8.3cm}|}
		\hline
		\textbf{No.} & \textbf{Problem} & \textbf{Drawbacks of TAI} & \textbf{Proposed GAI Approach}\\
		\hline
		\hline
		\cite{tolba2020massive} &CSI feedback &Cannot achieve performance as good as GAN  & Use GAN to recover original CSI from its compressed version.  \\ \cline{1-4}
		\cite{liang2020wireless} & CSI feedback & Cannot achieve performance as good as GAN  & Use GAN to enhance wireless channel data, resulting in better CSI compression processes \\ \cline{1-4}
		\cite{zhang2019generative} &Cell outage detection &Data imbalance issue  & Use GAN to generate more synthetic samples for minority classes  \\ \cline{1-4}
		\cite{vankayala2021radio} &Radio map estimation & Lack of training data & The generator aims to generate image masks while the discriminator learns to distinguish the masks of the original dataset and those generated by the generator \\ \cline{1-4}
		\cite{hussien2022prvnet} &CSI feedback & Less effective under noisy feedback channels  & Use VAE to compress CSI under noisy channel conditions  \\ \cline{1-4}
		\cite{zhang2023rme} &Radio map estimation &Poor performance due to nonuniformly positioned
		measurements and access constraints  & Use conditional GAN architecture to efficiently estimate radio maps based on observations from the environment  \\ \cline{1-4}
		\cite{xu2023ctgan} &Channel delay estimation & Lack of training data & Use GAN to generate synthetic cross-correlation data and smooth it with a Savitzky-Golay filter  \\ \hline\hline
		%		\cite{miuccio2022flexible} &Signal Encoding and decoding & Not mentioned & GAN & Use GAN to generate new codebooks  \\ \hline\hline
	\end{tabular}
	\label{tab:Summary_other_problems}
\end{table*}

\subsection{CSI Feedback}
With its powerful capabilities in learning data distribution and generating synthetic data, GAI has also been applied to recover compressed CSI feedback, as summarized in Table~\ref{tab:Summary_other_problems}. For example, the authors in~\cite{tolba2020massive} propose to use GAN for reconstructing CSI feedback in massive MIMO communications systems. In particular, massive MIMO can provide high cell throughput and reduce multiuser interference but largely relies on exploiting the CSI feedback from UEs. To reduce the signaling overhead of the system, the CSI feedback is usually compressed at UEs before transmitting to BSs. During the compressing process, important CSI information may be removed unintentionally, resulting in low precoding performance at BSs. To tackle this problem, the authors develop a GAN-based CSI recovery framework that can effectively generate a CSI matrix based on its compressed version. Specifically, the compressed CSI feedback will be first fed to the generator to estimate the CSI vector. This estimated CSI vector is then fed to the discriminator together with the original CSI vector to determine if the reconstructed CSI is good or bad. A new loss function combining the adversarial loss of the discriminator and the mean square error loss between the reconstructed and original CSI is introduced to further enhance the recovery performance of the proposed GAN-based approach. Extensive simulations reveal that by using GAN, the proposed framework is superior to traditional DL-based approaches. For instance, with a compression ratio of $\frac{1}{4}$, the GAN-based framework can achieve an outdoor NMSE of -15.88 dB while CsiNet~\cite{wen2018deep} and CsiNet+~\cite{guo2020convolutional} can only obtain -8.75 dB and -12.4 dB, respectively.

Differently, the authors in~\cite{hussien2022prvnet} propose to use VAEs for CSI compression at UEs under noisy channel conditions. The authors highlight that conventional DL-based CSI compression approaches like CsiNet in~\cite{wen2018deep} are vulnerable to noisy feedback channels which are common in practice. In contrast, the proposed VAE-based compressor can approximate distribution parameters for each dimension instead of estimating a point for each dimension in the latent space (i.e., deterministic latent space) as in classic DL-based solutions. As a result, the compressed CSI is robust against noise in the feedback channel. To make the proposed VAE network more suitable for the noise conditions of the feedback channel, the authors modify the VAE loss by using a weighted combination of reconstruction error and KL divergence between the encoder's distribution and the true distribution. The authors then test the proposed solution with an additive white Gaussian noise (AWGN) feedback channel and indicate that the proposed VAE-based compression technique can outperform other DL-based techniques (e.g., CsiNet~\cite{wen2018deep}) and compressive-sensing based models both under noise-free and noisy channel conditions. Similarly, the authors in~\cite{liang2020wireless} adopt GAN for wireless channel data augmentation before feeding CSI data into CsiNet. Specifically, a GAN-based network is developed to enrich data features of the original wireless channel data and also to generate new similar data by learning the distribution of the original channel data. These GAN-generated data will be fed into CsiNet for compression before feeding back to BSs. Simulation results reveal that using GAN can achieve a 3dB performance improvement compared to existing data augmentation approaches.

\subsection{Radio Map and Channel Delay Estimation}

Due to its capability of variational learning and sampling to explore the data distribution in a more versatile manner, GAI can also be used for radio map estimation~\cite{vankayala2021radio, zhang2023rme}, as summarized in Table~\ref{tab:Summary_other_problems}. In particular, a radio map spatially shows RF signal strength distribution and network coverage information which are essential characteristics for resource management and network planning in wireless communication systems. Unfortunately, conventional DL-based approaches such as RadioUNet~\cite{levie2021radiounet} and autoencoder~\cite{teganya2021deep} may not be effective for radio map estimation in modern IoT and cellular systems due to nonuniformly positioned measurements and access constraints. For that, the authors in~\cite{vankayala2021radio} and~\cite{zhang2023rme} propose to use the conditional GAN architecture to efficiently estimate radio maps based on observations from the environment. Particularly, the generator aims to generate image masks while the discriminator learns to distinguish the masks of the original dataset and those generated by the generator. Simulation results then demonstrate the effectiveness of GAN in estimating radio maps in various outdoor environments.

In addition, GAN is a promising approach for channel delay estimation as studied in~\cite{xu2023ctgan}. The authors aim to accurately estimate the first-arrival-path delay in wireless multi-path channels which plays an essential role in positioning and localization services. To do that, they first propose a CNN network to learn the mapping between the cross-correlation sequence and the delay offset. However, this CNN network suffers from the lack of training data. As such, the authors use GAN to generate synthetic cross-correlation data and smooth it with a Savitzky-Golay filter. The authors then perform various simulations to show that the proposed channel delay estimator can outperform existing approaches. In addition, the proposed GAN architecture can help to maintain a good estimation accuracy for the CNN network even with limited real cross-correlation data. Differently, the authors in~\cite{zhang2019generative} consider the cell outage detection problem in self-organizing cellular networks. Several classifiers based on DL have been proposed in the literature for this problem. However, as highlighted by the authors, these traditional approaches suffer from the data imbalance problem in which the number of training samples in one class is significantly larger than the number of samples in other classes. This leads to a biased classifier and degrades the service quality of the system. Therefore, the authors propose a novel GAN network with the aid of the Adaboost algorithm to preprocess the training data to change imbalanced data to balanced ones by generating more synthetic data for minority classes. Experimental results show that the proposed solution can effectively address the data imbalance problem and obtain better performance compared to state-of-the-art approaches.

\subsection{Summary and Lessons Learned}
With its capabilities in generating synthetic data under constraints, anomaly detection, uncertainty estimation, and variational learning and sampling, GAI has been widely adopted in the literature to address various problems such as physical layer security, channel estimation, signal classification, beamforming, JSCC, and IRS, as discussed in this section and summarized in Tables ~\ref{tab:Summary_modulation_signal_classification},~\ref{tab:Summary_channel_estimation_equalization},~\ref{tab:physical_layer_security},~\ref{tab:IRS},~\ref{tab:beamforming},~\ref{tab:JSCC}, and~\ref{tab:Summary_other_problems}. The lessons learned are as follows:

\begin{itemize}
	\item GAI has been mostly adopted for common issues in physical layer communications such as physical layer security, channel estimation, and signal detection. However, thanks to its capabilities, GAI can also be applied to other problems in the physical layer, opening new research directions. For example, the authors in~\cite{tolba2020massive} and~\cite{hussien2022prvnet} demonstrate that GAI is a great tool for efficient CSI compression. In addition, GAI can significantly improve the performance of JSCC and beamforming, as discussed in Sections~\ref{subsec:beamforming} and~\ref{subsec:JSCC}.
 
	\item The majority of GAI approaches in the literature are based on GAN due to its effectiveness in variational learning and sampling. Other approaches like VAEs, NFs, and diffusion models have been gaining more attention recently and are expected to be common in many applications in physical layer communications soon.
	
	\item Besides its great benefits, GAI can be used by adversaries to perform attacks at the physical layer as GAI can effectively generate fake data that is similar to real data from legitimate activities. However, research on countermeasures against GAI-based physical layer attacks is still limited, and more efforts from both academia and industry are required.
	
\end{itemize}

%\cite{tran2018generative, ferdowsi2019generative, tolba2020massive, liang2020wireless, shu2020generative, zhang2019generative, vankayala2021radio, yang2021radio, weisser2021generative, saarinen2021generative, xu2021wlan, kasgari2020experienced, he2022novel, marey2022pl, li2022variational, baur2022csi, hussien2022prvnet, chaker2022generative, zhang2023rme, wang2023radio, xu2023ctgan, miuccio2022flexible, zou2022underwater}
%=============================================
%=============================================
	\section{Open Issues and Future Research Directions}\label{sec:open_issues}
Although having great capabilities in complex data feature extraction, transformation, and enhancement, GAI is still in its early stage of development. Thus, open issues and research directions of GAI in physical layer communications will be discussed in this section.

\subsection{Security and Privacy}
As discussed above, adversarial attacks can significantly impact GAI systems. In particular, adversaries can inject crafted perturbations into the input data of GAI models to replicate these models or degrade their performance. Moreover, GAI can be exploited by adversaries to generate data that is similar to legitimate/trusted data, making conventional security approaches less effective in classifying these adversarial attacks. However, there is limited effort in dealing with adversarial attacks, especially GAI-based attacks in physical layer communications. One potential approach is to \textit{fight fire with fire} by using GAI models to generate adversarial training data and learn on this synthetic data to determine statistical anomalies that suggest potential perturbations. Moreover, GAI can be used to recover poisoned input data to mitigate the negative effects of adversarial perturbations~\cite{santhanam2018defending}.

\subsection{Model-driven GAI}
As can be observed in Section~\ref{sec:survey}, existing GAI-based models mostly focus on data-driven approaches that rely on the availability of training data. However, in practice, collecting a sufficient amount of training data may be costly, time-consuming, and even impossible. To tackle this issue, model-driven approaches~\cite{he2019model, he2020model} can be adopted. In particular, model-driven approaches can incorporate the prior knowledge of target domains, e.g., carrier frequencies, physical constraints, and noise distributions, into the training process to further improve the performance of GAI-based solutions. For example, with prior knowledge of bandwidth and carrier frequency, GAI-based solutions can be trained to generate more realistic channel samples.

\subsection{Resource-Efficient Learning}
The training and inference of GAI require computation, storage, and communication resources, putting burdens on existing communication systems, especially for resource-constrained devices such as IoT devices, mobile phones, and UAVs. As such, novel GAI architectures need to be developed to minimize resource consumption while maintaining good learning performance. Distributed and federated learning can be integrated into GAI to offload computational tasks to edge devices as well as reduce communication overhead by transmitting model updates instead of raw data. For example, GAI models can be trained at edge devices with local data and then aggregated at a centralized server to obtain a global GAI model. In addition, GAI can be used to recover compressed local model updates to reduce communication overhead while still maintaining good training performance. Incentivization mechanisms such as dynamic spectrum access should also be considered to utilize communication resources, especially in cognitive radio networks as studied in a few papers reviewed in Section~\ref{sec:survey}.

\subsection{Real-time Adaptation}
Although GAI has the capability of domain adaptation that can leverage knowledge from a source domain for training in a target domain, it still requires a large amount of training data and a long training time to achieve good performance. Consequently, GAI may not effectively deal with real-time wireless channel/environment changes caused by random factors such as mobility, blockage, and interference. For that, it is essential to develop novel GAI approaches that can quickly adapt to track these variations. Integrating advanced ML techniques like meta-learning~\cite{hospedales2021meta} into GAI is a promising direction to help it quickly adapt to new environmental conditions based on a few training samples. Specifically, meta-learning can obtain important and useful information in the training process of source environments and use that knowledge to quickly learn new environments. With meta-learning, GAI can obtain good accuracy with a few training data samples in new wireless systems, making it more practical in real-world applications. In addition, over-the-air evaluation and implicit CSI feedback mechanisms should be developed to further improve the performance of GAI under the dynamics and uncertainty of physical layer communications.

%=============================================
%=============================================
	%-----------------------------------------------------------------------------------------------------------
	
	\section{Conclusion}\label{sec:Summary}
	Generative AI is a promising technology for physical layer communications due to its capabilities of complex data feature extraction, transformation, and enhancement. In this article, we have presented a comprehensive survey of the applications of generative AI in physical layer communications. Firstly, we have introduced an overview of generative AI, common generative models, and their advantages compared to traditional AI techniques. Then, we have provided detailed reviews, analyses, and comparisons of different generative AI techniques in emerging problems in physical layer communications such as channel modeling, channel estimation and signal detection, physical layer security, joint source channel coding, beamforming, and intelligent reflecting surface. Finally, we have highlighted important open issues and future research directions of generative AI in physical layer communications.
	%%-----------------------------------------------------------------------------------------------------------
	
	\bibliographystyle{ieeetran}
	
	\bibliography{Bibtex/introduction,Bibtex/Rwork,Bibtex/survey,Bibtex/open_issues}
	
\end{document}